\documentclass{svmult}

\DeclareMathAlphabet{\mathcal}{OMS}{cmsy}{m}{n}

\usepackage{amsfonts}
\usepackage{amsmath}
\usepackage{graphicx}
\usepackage[ruled,vlined]{algorithm2e}
\usepackage{verbatim}
\usepackage{nccmath}
\usepackage{url}
\usepackage{mathptmx}       
\usepackage{helvet}         
\usepackage{courier}        
\usepackage{type1cm}        
\usepackage{makeidx}         
\usepackage{graphicx}        
\usepackage{multicol}        
\usepackage{multirow}
\usepackage[bottom]{footmisc}
\usepackage{array}

\makeindex             

\usepackage{amsfonts}
\usepackage{amsmath}
\usepackage{amssymb}
\usepackage{appendix}
\usepackage{array}
\usepackage{bbold}
\usepackage{bigstrut}
\usepackage{bm}
\usepackage[bottom]{footmisc}
\usepackage{caption}
\usepackage{color}
\usepackage{courier}
\usepackage{float}
\usepackage{george}
\usepackage{graphicx}
\usepackage{graphics}
\usepackage{helvet}
\usepackage{hyperref}
\usepackage{longtable}
\usepackage{makeidx}
\usepackage{multicol}
\usepackage{multirow}
\usepackage{nccmath}
\usepackage{pgf}
\usepackage[ruled,vlined]{algorithm2e}
\usepackage{santosh}
\usepackage{setspace}
\usepackage{soul}
\usepackage{subfigure}
\usepackage{threeparttable}
\usepackage{tikz}
\usepackage{times}
\usepackage{type1cm}
\usepackage{url}
\usepackage[varg]{txfonts}
\usepackage{verbatim}
\usepackage{xspace}

\newcommand{\SLIM}{{$\mathtt{SLIM}$}\xspace}
\newcommand{\fsSLIM}{{$\mathtt{fsSLIM}$}\xspace}

\newcommand{\GLSLIM}{$\mathtt{GLSLIM}$\xspace}
\newcommand{\recwalkKstep}{$\mathtt{RecWalk}^\mathtt{K-step}$\xspace}
\newcommand{\recwalkPR}{$\mathtt{RecWalk}^\mathtt{PR}$\xspace}
\newcommand{\recwalk}{$\mathtt{RecWalk}$\xspace}
\newcommand{\perdif}{{$\mathtt{PerDif}$}\xspace}
\newcommand{\dict}{\textsc{PerDif\textsuperscript{par}}\xspace}
\newcommand{\free}{\textsc{PerDif\textsuperscript{free}}\xspace}

\newcommand{\ol}[1]{\overline{#1}}
\newcommand{\tr}[1]{#1^\top}
\newcommand{\UU}{\mathcal{U}}
\newcommand{\II}{\mathcal{I}}
\newcommand{\RR}{\mathcal{R}}
\newcommand{\SSS}{\mathcal{S}}
\newcommand{\NN}{\mathcal{N}}
\newcommand{\PP}{\mathrm{Pr}}
\newcommand{\ppi}{\vec \piup}
\newcommand{\pp}{\vec p}
\newcommand{\qq}{\vec q}
\newcommand{\sss}{\vec s}
\newcommand{\eee}{\vec e}
\newcommand{\dd}{\vec d}

\newcommand{\hh}{\vec h}
\newcommand{\xx}{\vec x}
\newcommand{\rr}{\vec r}

\newcommand{\ww}{\vec w}
\newcommand{\bb}{\vec b}

\newcommand{\ssum}{\sum\limits}
\newcommand{\msum}{\sum}
\DeclareMathOperator*{\argmax}{arg\,max}

\newcolumntype{x}[1]{
  >{\centering\hspace{0pt}}p{#1}}%
\newcommand{\tn}{\tabularnewline}

\newcommand{\itempar}{\hspace{\svparindent}}

\newcommand{\beq}{\begin{equation}}
\newcommand{\eeq}{\end{equation}}
\newcommand{\beqa}{\begin{eqnarray}}
\newcommand{\eeqa}{\end{eqnarray}}
\newcommand{\beqas}{\begin{eqnarray*}}
\newcommand{\eeqas}{\end{eqnarray*}}
\usepackage[notheorems]{nikolako-math}

\definecolor{Maroon}{RGB}{238,83,34}
\definecolor{Maroon}{RGB}{25,89,121}
\definecolor{bgorange}{RGB}{238,83,0}
\definecolor{bgorangelight}{RGB}{255,173,115}
\definecolor{bggreen}{RGB}{237,255,178}
\definecolor{bgblue}{RGB}{128,255,204}
\definecolor{Maroon}{RGB}{122,0,25}
\definecolor{Gold}{RGB}{255,204,51} 

\definecolor{mygreen}{rgb}{0.576235910262533,0.6647921100239956,0.0}
\definecolor{myorange}{rgb}{0.8273618134500182,0.5740127684739248,0.0}
\definecolor{myblue}{rgb}{0.3804515572007965,0.6115286754014934,1.0}

\usepackage{pgfplots}
\pgfplotsset{compat=newest}
\usepgfplotslibrary{groupplots}
\usepgfplotslibrary{polar}
\usepgfplotslibrary{statistics}
\usetikzlibrary{positioning,chains,fit,shapes,calc}
\usetikzlibrary{shapes}
\definecolor{pinegreen}{cmyk}{0.92,0,0.59,0.25}
\definecolor{royalblue}{cmyk}{1,0.50,0,0}
\definecolor{lavander}{cmyk}{0,0.48,0,0}
\definecolor{violet}{cmyk}{0.79,0.88,0,0}
\tikzstyle{citems}=[circle, draw, thin,fill=Gold, scale=0.8]
\tikzstyle{cusers}=[rectangle, draw, thin,fill=Maroon, scale=0.8]
\tikzstyle{cusers2}=[rectangle, draw, thin,fill=white, scale=0.8]
\tikzstyle{cred}=[circle, draw, thin,fill=Maroon, scale=0.8]
\tikzstyle{cgreen}=[circle, draw, thin,fill=royalblue, scale=0.8]
\tikzstyle{rpath}=[ultra thick, Maroon, opacity=0.8]
\tikzstyle{gpath}=[ultra thick, royalblue, opacity=0.8]
\makeatletter
\global\let\tikz@ensure@dollar@catcode=\relax
\makeatother

\begin{document}


\title*{Trust your neighbors: A comprehensive survey of neighborhood-based methods for recommender systems}

\author{Athanasios N. Nikolakopoulos, Xia Ning, Christian Desrosiers and George Karypis}
\institute{Athanasios N. Nikolakopoulos \at Amazon, Seattle, USA\\
\email{annikolako@gmail.com}\\ 
(work done prior to joining Amazon)
\and 
Xia Ning \at Biomedical Informatics, Computer Science and Engineering dep., The Ohio State University, Columbus, USA \\ \email{ning.104@osu.edu}
\and Christian Desrosiers \at Software Engineering and IT dep., \'Ecole de Technologie Sup\'erieure, Montreal, Canada \\ \email{christian.desrosiers@etsmtl.ca}
\and George Karypis \at Computer Science \& Engineering dep., University of Minnesota, Minneapolis, USA \\ \email{karypis@cs.umn.edu}}
%
%
\maketitle

\abstract{
Collaborative recommendation approaches based on nearest-neighbors are still highly popular today due to their simplicity, their efficiency, and their ability to produce accurate and personalized recommendations. This chapter offers a comprehensive survey of neighborhood-based methods for the item recommendation problem. It presents the main characteristics and benefits of such methods, describes key design choices for implementing a neighborhood-based recommender system, and gives practical information on how to make these choices. 
A broad range of methods is covered in the chapter, including traditional algorithms like k-nearest neighbors as well as advanced approaches based on matrix factorization, sparse coding and random walks.
}


\section{Introduction}\label{sec:introduction}

The appearance and growth of online markets has had a considerable impact on the habits of consumers, providing them access to a greater variety of products and information on these goods. While this freedom of purchase has made online commerce into a multi-billion dollar industry, it also made it more difficult for consumers to select the products that best fit their needs. One of the main solutions proposed for this information overload problem are recommender systems, which provide automated and personalized suggestions of products to consumers. 

The recommendation problem can be defined as estimating the response of a user for unseen items, based on historical information stored in the system, and suggesting to this user \emph{novel} and \emph{original} items for which the predicted response is \emph{high}. User-item responses can be numerical values known as ratings (e.g., 1-5 stars), ordinal values (e.g., strongly agree, agree, neutral, disagree, strongly disagree) representing the possible levels of user appreciation, or binary values (e.g., like/dislike or interested/not interested). Moreover, user responses can be obtained explicitly, for instance, through ratings/reviews entered by users in the system, or implicitly, from purchase history or access patterns \cite{konstan97,terveen97}. For the purpose of simplicity, from this point on, we will call rating any type of user-item response.

Item recommendation \index{item recommendation} approaches can be divided in two broad categories: personalized and non-personalized. Among the personalized approaches are \emph{content-based} and \emph{collaborative filtering} methods, as well as \emph{hybrid} techniques combining these two types of methods. 
The general principle of con\-tent-based (or cognitive) methods \cite{balabanovic97,billsus00,lang95,pazzani97} is to identify the common characteristics of items that have received a favorable rating from a user, and then recommend to this user unseen items that share these characteristics. Recommender systems \index{recommender systems} based purely on content generally suffer from the problems of \emph{limited content analysis} and \emph{over-specialization} \cite{shardanand95}. Limited content analysis occurs when the system has a limited amount of information on its users or the content of its items. For instance, privacy issues might refrain a user from providing personal information, or the precise content of items may be difficult or costly to obtain for some types of items, such as music or images. Another problem is that the content of an item is often insufficient to determine its quality. Over-specialization, on the other hand, is a side effect of the way in which content-based systems recommend unseen items, where the predicted rating of a user for an item is high if this item is similar to the ones liked by this user. For example, in a movie recommendation application, the system may recommend to a user a movie of the same genre or having the same actors as movies already seen by this user. Because of this, the system may fail to recommend items that are different but still interesting to the user. More information on content-based recommendation approaches can be found in Chapter~\ref{24-content-rs} of this book.

Instead of depending on content information, collaborative (or social) filtering approaches use the rating information of other users and items in the system. The key idea is that the rating of a target user for an unseen item is likely to be similar to that of another user, if both users have rated other items in a similar way. Likewise, the target user is likely to rate two items in a similar fashion, if other users have given similar ratings to these two items. Collaborative filtering approaches overcome some of the limitations of content-based ones. For instance, items for which the content is not available or difficult to obtain can still be recommended to users through the feedback of other users. Furthermore, collaborative recommendations are based on the quality of items as evaluated by peers, instead of relying on content that may be a bad indicator of quality. Finally, unlike content-based systems, collaborative filtering ones can recommend items with very different content, as long as other users have already shown interest for these different items.

Collaborative filtering \index{collaborative filtering} approaches can be grouped in two general classes of \emph{neighborhood} and \emph{model}-based methods. In neighborhood-based (memory-based \cite{breese98} or heuristic-based \cite{adomavicius05}) collaborative filtering \cite{delgado99,deshpande04,hill95,konstan97,linden03,nakamura98,resnick94,sarwar01,shardanand95}, the user-item ratings stored in the system are directly used to predict ratings for unseen items. This can be done in two ways known as \emph{user-based} or \emph{item-based} recommendation. User-based systems, such as GroupLens \cite{konstan97}, 
evaluate the interest of a target user for an item using the ratings for this item by other users, called \emph{neighbors}, that have similar rating patterns. The neighbors of the target user are typically the users whose ratings are most correlated to the target user's ratings. Item-based approaches \cite{deshpande04,linden03,sarwar01}, on the other hand, predict the rating of a user for an item based on the ratings of the user for similar items. In such approaches, two items are similar if several users of the system have rated these items in a similar fashion. 

In contrast to neighborhood-based systems, which use the stored ratings directly in the prediction, model-based approaches use these ratings to learn a predictive model. Salient characteristics of users and items are captured by a set of model parameters, which are learned from training data and later used to predict new ratings. Model-based approaches for the task of recommending items are numerous and include Bayesian Clustering \cite{breese98}, Latent Semantic Analysis \cite{hofmann03}, Latent Dirichlet Allocation \cite{blei03}, Maximum Entropy \cite{zitnick04}, Boltzmann Machines \cite{salakhutdinov07}, Support Vector Machines \cite{grcar06}, and Singular Value Decomposition \cite{bell07a,koren08,paterek07,takacs08,takacs09}. A survey of state-of-the-art model-based methods can be found in Chapter \ref{15-collab-filt-rs} of this book. 

Finally, to overcome certain limitations of content-based and collaborative filtering methods, hybrid recommendation approaches combine characteristics of both types of methods. Content-based and collaborative filtering methods can be combined in various ways, for instance, by merging their individual predictions into a single, more robust prediction \cite{pazzani99,billsus00}, or by adding content information into a collaborative filtering model \cite{Adams2010,Agarwal2011,Yoo2009,Singh2008,nikolakopoulos2015hierarchical,r14,nikolakopoulos2015top}. Several studies have shown hybrid recommendation approaches to provide more accurate recommendations than pure content-based or collaborative methods, especially when few ratings are available \cite{adomavicius05}.

\subsection{Advantages of neighborhood approaches}

While recent investigations show state-of-the-art model-based approaches superior to neighborhood ones in the task of predicting ratings \cite{koren08,takacs07}, there is also an emerging understanding that good prediction accuracy alone does not guarantee users an effective and satisfying experience \cite{herlocker04}. Another factor that has been identified as playing an important role in the appreciation of users for the recommender system is \emph{serendipity} \cite{herlocker04,sarwar01}. Serendipity extends the concept of novelty by helping a user find an interesting item he or she might not have otherwise discovered. For example, recommending to a user a movie directed by his favorite director constitutes a novel recommendation if the user was not aware of that movie, but is likely not serendipitous since the user would have discovered that movie on his own. A more detailed discussion on novelty and diversity is provided in Chapter~\ref{5-novelty} of this book.

Model-based approaches excel at characterizing the preferences of a user with latent factors. For example, in a movie recommender system, such methods may determine that a given user is a fan of movies that are both funny and romantic, without having to actually define the notions ``funny'' and ``romantic''. This system would be able to recommend to the user a romantic comedy that may not have been known to this user. However, it may be difficult for this system to recommend a movie that does not quite fit this high-level genre, for instance, a funny parody of horror movies. Neighborhood approaches, on the other hand, capture local associations in the data. Consequently, it is possible for a movie recommender system based on this type of approach to recommend the user a movie very different from his usual taste or a movie that is not well known (e.g., repertoire film), if one of his closest neighbors has given it a strong rating. This recommendation may not be a guaranteed success, as would be a romantic comedy, but it may help the user discover a whole new genre or a new favorite actor/director.  

The main advantages of neighborhood-based methods are:
\begin{itemize}

\item \textbf{Simplicity:} Neighborhood-based methods are intuitive and relatively simple to implement. In their simplest form, only one parameter (the number of neighbors used in the prediction) requires tuning.

\vspace{2mm}
 
\item \textbf{Justifiability:} Such methods also provide a concise and intuitive justification for the computed predictions. For example, in item-based recommendation, the list of neighbor items, as well as the ratings given by the user to these items, can be presented to the user as a justification for the recommendation. This can help the user better understand the recommendation and its relevance, and could serve as basis for an interactive system where users can select the neighbors for which a greater importance should be given in the recommendation \cite{bell07a}. The benefits and challenges of explaining recommendations to users are addressed in Chapter~\ref{8-explanations} of this book.

\vspace{2mm}

\item \textbf{Efficiency:} One of the strong points of neighborhood-based systems are their efficiency. Unlike most model-based systems, they require no costly training phases, which need to be carried at frequent intervals in large commercial applications. 
These systems may require pre-computing nearest neighbors in an offline step, which is typically much cheaper than model training, 
providing near instantaneous recommendations. Moreover, storing these nearest neighbors requires very little memory, making such approaches scalable to applications having millions of users and items.   

\vspace{2mm}

\item \textbf{Stability:} Another useful property of recommender systems based on this approach is that they are little affected by the constant addition of users, items and ratings, which are typically observed in large commercial applications. For instance, once item similarities have been computed, an item-based system can readily make recommendations to new users, without having to re-train the system. Moreover, once a few ratings have been entered for a new item, only the similarities between this item and the ones already in the system need to be computed.

\end{itemize}

While neighborhood-based methods have gained popularity due to these advantages\footnote{For further insights of some of the key properties of neighborhood-based methods under a probabilistic lens,  see \cite{canamares2017probabilistic}. Therein, the interested reader can find a probabilistic reformulation of basic neighborhood-based methods that elucidates certain aspects of their effectiveness; delineates innate connections with item popularity; while also, allows for comparisons between basic  neighborhood-based variants.}, they are also known to suffer from the problem of limited coverage, which causes some items to be never recommended. Also, traditional methods of this category are known to be more sensitive to the sparseness of ratings and the cold-start problem, where the system has only a few ratings, or no rating at all, for new users and items. Section \ref{sec:advanced-techniques} presents more advanced neighborhood-based techniques that can overcome these problems.

\subsection{Objectives and outline}

This chapter has two main objectives. It first serves as a general guide on neighborhood-based recommender systems, and presents practical information on how to implement such recommendation approaches. In particular, the main components of neighborhood-based methods will be described, as well as the benefits of the most common choices for each of these components. Secondly, it presents more specialized techniques on the subject that address particular aspects of recommending items, such as data sparsity. Although such techniques are not required to implement a simple neighborhood-based system, having a broader view of the various difficulties and solutions for neighborhood methods may help making appropriate decisions during the implementation process.

The rest of this document is structured as follows. In Section \ref{sec:definition-notation}, we first give a formal definition of the item recommendation task and present the notation used throughout the chapter. In Section \ref{sec:nb-recommendation}, the principal neighborhood approaches, predicting user ratings for unseen items based on regression or classification, are then introduced, and the main advantages and flaws of these approaches are described. This section also presents two complementary ways of implementing such approaches, either based on user or item similarities, and analyzes the impact of these two implementations on the accuracy, efficiency, stability, justfiability ans serendipity of the recommender system. Section \ref{sec:components-nb-methods}, on the other hand, focuses on the three main components of neighborhood-based recommendation methods: rating normalization, similarity weight computation, and neighborhood selection. For each of these components, the most common approaches are described, and their respective benefits compared. In Section \ref{sec:advanced-techniques}, the problems of limited coverage and data sparsity are introduced, and several solutions proposed to overcome these problems are described. In particular, several techniques based on dimensionality reduction and graphs are presented. Finally, the last section of this document summarizes the principal characteristics and methods of neighorhood-based recommendation, and gives a few more pointers on implementing such methods.

\section{Problem definition and notation}\label{sec:definition-notation}

In order to give a formal definition of the item recommendation task, we introduce the following notation. The set of users in the recommender system will be denoted by $\UU$, and the set of items by $\II$. Moreover, we denote by $\RR$ the set of ratings recorded in the system, and write $\SSS$ the set of possible values for a rating (e.g., $\SSS = [1,5]$ or $\SSS = \{\textrm{like},\textrm{dislike}\}$). Also, we suppose that no more than one rating can be made by any user $u \in \UU$ for a particular item $i \in \II$ and write $r_{ui}$ this rating. To identify the subset of users that have rated an item $i$, we use the notation $\UU_i$. Likewise, $\II_u$ represents the subset of items that have been rated by a user $u$. Finally, the items that have been rated by two users $u$ and $v$, i.e. $\II_u \cap \II_v$, is an important concept in our presentation, and we use $\II_{uv}$ to denote this concept. In a similar fashion, $\UU_{ij}$ is used to denote the set of users that have rated both items $i$ and $j$.

Two of the most important problems associated with recommender systems are the \emph{rating prediction}
and \emph{top-$N$} recommendation problems. 
The first problem is to predict the rating that a user $u$ will give his or  her unrated item $i$. 
When ratings are available, this task is most often defined as a regression or (multi-class) classification problem where the goal is to learn a function 
$f : \UU \times \II \to \SSS$ that predicts the rating $f(u,i)$ of a user $u$ for an unseen item $i$. 
Accuracy is commonly used to evaluate the performance of the recommendation method. Typically, the ratings $\RR$ are divided into a \emph{training} set $\RR_\text{train}$ used to learn $f$, and a \emph{test} set $\RR_\text{test}$ used to evaluate the prediction accuracy. Two popular measures of accuracy are the \emph{Mean Absolute Error} (MAE):
\beq
  \text{MAE}(f) \, = \, \frac{1}{|\RR_\text{test}|} 
        \msum_{r_{ui} \in \RR_\text{test}} \!\!|f(u,i) - r_{ui}|, 
\eeq
and the \emph{Root Mean Squared Error} (RMSE):
\beq
  \text{RMSE}(f) \, = \, \sqrt{\frac{1}{|\RR_\text{test}|} 
            \msum_{r_{ui} \in \RR_\text{test}} \!\!\left(f(u,i) - r_{ui}\right)^2}.
\eeq
When ratings are not available, for instance, if only the list of items purchased by each user is known, measuring the rating prediction accuracy is not possible. In such cases, the problem of finding the best item is usually transformed into the task of recommending to an active user $u_a$ a list $L(u_a)$ containing $N$ items likely to interest him or her \cite{deshpande04,sarwar01}. The quality of such method can be evaluated by splitting the items of $\II$ into a set $\II_\text{train}$, used to learn $L$, and a test set $\II_\text{test}$. Let $T(u) \subset \II_u \cap \II_\text{test}$ be the subset of test items that a user $u$ found relevant. If the user responses are binary, these can be the items that $u$ has rated positively. Otherwise, if only a list of purchased or accessed items is given for each user $u$, then these items can be used as $T(u)$. The performance of the method is then computed using the measures of \emph{precision} and \emph{recall}:
\begin{eqnarray}
 \text{Precision}(L) & \, = \, & \frac{1}{|\UU|} \msum_{u \in \UU} |L(u) \cap T(u)| \, / \, |L(u)| \\
 \text{Recall}(L) & \, = \, & \frac{1}{|\UU|} \msum_{u \in \UU} |L(u) \cap T(u)| \, / \, |T(u)|.
\end{eqnarray}
A drawback of this task is that all items of a recommendation list $L(u)$ are considered equally interesting to user $u$. An alternative setting, described in \cite{deshpande04}, consists in learning a function $L$ that maps each user $u$ to a list $L(u)$ where items are \emph{ordered} by their ``interestingness'' to $u$. If the test set is built by randomly selecting, for each user $u$, a single item $i_u$ of $\II_u$, the performance of $L$ can be evaluated with the \emph{Average Reciprocal Hit-Rank} (ARHR):
\beq
 \text{ARHR}(L) \, = \, \frac{1}{|\UU|} \msum_{u \in \UU}  \frac{1}{\text{rank}(i_u, L(u))},
\eeq
where $\text{rank}(i_u, L(u))$ is the rank of item $i_u$ in $L(u)$, equal to $\infty$ if $i_u \not\in L(u)$. A more extensive description of evaluation measures for recommender systems can be found in Chapter~\ref{29-evaluation} of this book.

\section{Neighborhood-based recommendation}\label{sec:nb-recommendation}\index{neighborhood-based recommendation}

Recommender systems based on neighborhood automate the common principle that similar users prefer similar items, and similar items are preferred by similar users. 
%
To illustrate this, consider the following example based on the ratings of Figure \ref{fig:toy-example}.

\begin{example}
User Eric has to decide whether or not to rent the movie ``Titanic'' that he has not yet seen. He knows that Lucy has very similar tastes when it comes to movies, as both of them hated ``The Matrix'' and loved ``Forrest Gump,'' so he asks her opinion on this movie. On the other hand, Eric finds out he and Diane have different tastes, Diane likes action movies while he does not, and he discards her opinion or considers the opposite in his decision. 
\end{example}

\begin{figure}[h!tb]
\begin{center}
\begin{small}
\begin{tabular}{x{1cm}||x{1cm}|x{1cm}|x{1cm}|x{1cm}|x{1cm}}
\hline
\multirow{2}{*}{} & The    & \multirow{2}{*}{Titanic} & Die  & Forrest & \multirow{2}{*}{Wall-E}\tn
                  & Matrix &                          & Hard & Gump    & \tn
\hline\hline
John   &   5   &   1   &       &   2   &  2  \tn
Lucy   &   1   &   5   &   2   &   5   &  5  \tn
Eric   &   2   &   ?   &   3   &   5   &  4  \tn
Diane  &   4   &   3   &   5   &   3   &     \tn
\hline
\end{tabular}
\end{small}
\caption{A ``toy example'' showing the ratings of four users for five movies.}
\label{fig:toy-example}
\end{center}
\end{figure}

\subsection{User-based rating prediction}

User-based neighborhood recommendation methods predict the rating $r_{ui}$ of a user $u$ for an unseen item $i$ using the ratings given to $i$ by users most similar to $u$, called nearest-neighbors. Suppose we have for each user $v \neq u$ a value $w_{uv}$ representing the preference similarity between $u$ and $v$ (how this similarity can be computed will be discussed in Section \ref{sec:sim-weight-computation}). The $k$-nearest-neighbors ($k$-NN) of $u$, denoted by $\NN(u)$, are the $k$ users $v$ with the highest similarity $w_{uv}$ to $u$. However, only the users who have rated item $i$ can be used in the prediction of $r_{ui}$, and we instead consider the $k$ users most similar to $u$ that \emph{have rated} $i$. We write this set of neighbors as $\NN_i(u)$. The rating $r_{ui}$ can be estimated as the average rating given to $i$ by these neighbors:
\beq\label{eqn:simple-ub-prediction}
 \hat{r}_{ui} \, = \, \frac{1}{|\NN_i(u)|} \msum_{v \in \NN_i(u)}\!\!\! r_{vi}.
\eeq
A problem with (\ref{eqn:simple-ub-prediction}) is that is does not take into account the fact that the neighbors can have different levels of similarity. Consider once more the example of Figure~\ref{fig:toy-example}. If the two nearest-neighbors of Eric are Lucy and Diane, it would be foolish to consider equally their ratings of the movie ``Titanic,'' since Lucy's tastes are much closer to Eric's than Diane's. A common solution to this problem is to weigh the contribution of each neighbor by its similarity to $u$. However, if these weights do not sum to $1$, the predicted ratings can be well outside the range of allowed values. Consequently, it is customary to normalize these weights, such that the predicted rating becomes
\beq\label{eqn:unnormalized-ub-prediction}
 \hat{r}_{ui} \, = \, \frac{\ssum_{v \in \NN_i(u)}\!\!\! w_{uv} \, r_{vi}}
                    {\ssum_{v \in \NN_i(u)}\!\!\! |w_{uv}|}.
\eeq
In the denominator of (\ref{eqn:unnormalized-ub-prediction}), $|w_{uv}|$ is used instead of $w_{uv}$ because negative weights can produce ratings outside the allowed range. Also, $w_{uv}$ can be replaced by $w^\alpha_{uv}$, where $\alpha > 0$ is an amplification factor \cite{breese98}. When $\alpha > 1$, as is it most often employed, an even greater importance is given to the neighbors that are the closest to $u$. 

\begin{example}
Suppose we want to use (\ref{eqn:unnormalized-ub-prediction}) to predict Eric's rating of the movie ``Titanic'' using the ratings of Lucy and Diane for this movie. Moreover, suppose the similarity weights between these neighbors and Eric are respectively $0.75$ and $0.15$. The predicted rating would be
\begin{displaymath}
\hat{r} \, = \, \frac{0.75\!\times\!5 \, + \,0.15\!\times\!3}{0.75 \, + \, 0.15} \ \simeq \ 4.67, 
\end{displaymath}
which is closer to Lucy's rating than to Diane's.
\end{example}

Equation (\ref{eqn:unnormalized-ub-prediction}) also has an important flaw: it does not consider the fact that users may use different rating values to quantify the same level of appreciation for an item. For example, one user may give the highest rating value to only a few outstanding items, while a less difficult one may give this value to most of the items he likes. This problem is usually addressed by converting the neighbors' ratings $r_{vi}$ to normalized ones $h(r_{vi})$ \cite{breese98,resnick94}, giving the following prediction:
\beq\label{eqn:normalized-ub-prediction}
 \hat{r}_{ui} \, = \, h^{-1}\left(\frac{\ssum_{v \in \NN_i(u)}\!\!\! w_{uv} \, h(r_{vi})}
                    {\ssum_{v \in \NN_i(u)}\!\!\! |w_{uv}|}\right).
\eeq
Note that the predicted rating must be converted back to the original scale, hence the $h^{-1}$ in the equation. The most common approaches to normalize ratings will be presented in Section \ref{sec:rating-normalization}.

\subsection{User-based classification}

The prediction approach just described, where the predicted ratings are computed as a weighted average of the neighbors' ratings, essentially solves a \emph{regression} problem. Neighborhood-based \emph{classification}, on the other hand, finds the most likely rating given by a user $u$ to an item $i$, by having the nearest-neighbors of $u$ vote on this value. The vote $v_{ir}$ given by the $k$-NN of $u$ for the rating $r \in \SSS$ can be obtained as the sum of the similarity weights of neighbors that have given this rating to $i$: 
\beq\label{eqn:unnormalized-ub-classification}
  v_{ir} \, = \, \msum_{v \in \NN_i(u)}\!\!\! \delta(r_{vi} = r) \, w_{uv},
\eeq
where $\delta(r_{vi} = r)$ is $1$ if $r_{vi} = r$, and $0$ otherwise. Once this has been computed for every possible rating value, the predicted rating is simply the value $r$ for which $v_{ir}$ is the greatest. 

\begin{example}
Suppose once again that the two nearest-neighbors of Eric are Lucy and Diane with respective similarity weights $0.75$ and $0.15$. In this case, ratings $5$ and $3$ each have one vote. However, since Lucy's vote has a greater weight than Diane's, the predicted rating will be $\hat{r} = 5$.  
\end{example}

A classification method that considers normalized ratings can also be defined. Let $\SSS'$ be the set of possible normalized values (that may require discretization), the predicted rating is obtained as:
\beq\label{eqn:normalized-ub-classification}
  \hat{r}_{ui} \, = \, h^{-1}\left( \argmax_{r \in \SSS'} \, \msum_{v \in \NN_i(u)}\!\!\! \delta(h(r_{vi}) = r) \, w_{uv} \right).
\eeq

\subsection{Regression VS classification}

The choice between implementing a neighborhood-based regression or classification method largely depends on the system's rating scale. Thus, if the rating scale is continuous, e.g. ratings in the \emph{Jester} joke recommender system \cite{goldberg01} can take any value between $-10$ and $10$, then a regression method is more appropriate. On the contrary, if the rating scale has only a few discrete values, e.g. ``good'' or ``bad,'' or if the values cannot be ordered in an obvious fashion, then a classification method might be preferable. Furthermore, since normalization tends to map ratings to a continuous scale, it may be harder to handle in a classification approach.

Another way to compare these two approaches is by considering the situation where all neighbors have the same similarity weight. As the number of neighbors used in the prediction increases, the rating $r_{ui}$ predicted by the regression approach will tend toward the mean rating of item $i$. Suppose item $i$ has only ratings at either end of the rating range, i.e. it is either loved or hated, then the regression approach will make the safe decision that the item's worth is average. This is also justified from a statistical point of view since the expected rating (estimated in this case) is the one that minimizes the RMSE. On the other hand, the classification approach will predict the rating as the most frequent one given to $i$. This is more risky as the item will be labeled as either ``good'' or ``bad''. However, as mentioned before, risk taking may be be desirable if it leads to serendipitous recommendations.

\subsection{Item-based recommendation}\index{item-based recommendation}

While user-based methods rely on the opinion of like-minded users to predict a rating, item-based approaches \cite{deshpande04,linden03,sarwar01} look at ratings given to similar items. Let us illustrate this approach with our toy example. 

\begin{example}
Instead of consulting with his peers, Eric instead determines whether the movie ``Titanic'' is right for him by considering the movies that he has already seen. He notices that people that have rated this movie have given similar ratings to the movies ``Forrest Gump'' and ``Wall-E''. Since Eric liked these two movies he concludes that he will also like the movie ``Titanic''.
\end{example}

This idea can be formalized as follows. Denote by $\NN_u(i)$ the items rated by user $u$ most similar to item $i$. The predicted rating of $u$ for $i$ is obtained as a weighted average of the ratings given by $u$ to the items of $\NN_u(i)$: 
\beq\label{eqn:unnormalized-ib-prediction}
 \hat{r}_{ui} \, = \, \frac{\ssum_{j \in \NN_u(i)}\!\!\! w_{ij} \, r_{uj}}
                    {\ssum_{j \in \NN_u(i)}\!\!\! |w_{ij}|}.
\eeq

\begin{example}
Suppose our prediction is again made using two nearest-neighbors, and that the items most similar to ``Titanic'' are ``Forrest Gump'' and ``Wall-E,'' with respective similarity weights $0.85$ and $0.75$. Since ratings of $5$ and $4$ were given by Eric to these two movies, the predicted rating is computed as 
\begin{displaymath}
  \hat{r} \, = \, \frac{0.85\!\times\!5 \, + \,0.75\!\times\!4}{0.85 \, + \, 0.75} \ \simeq \ 4.53.
\end{displaymath}
\end{example}

Again, the differences in the users' individual rating scales can be considered by normalizing ratings with a $h$:
\beq\label{eqn:normalized-ib-prediction}
 \hat{r}_{ui} \, = \, h^{-1}\left(\frac{\ssum_{j \in \NN_u(i)}\!\!\! w_{ij} \, h(r_{uj})}
                    {\ssum_{j \in \NN_u(i)}\!\!\! |w_{ij}|}\right).
\eeq
Moreover, we can also define an item-based classification approach. In this case, the items $j$ rated by user $u$ vote for the rating to be given to an unseen item $i$, and these votes are weighted by the similarity between $i$ and $j$. The normalized version of this approach can be expressed as follows:
\beq\label{eqn:normalized-ib-classification}
  \hat{r}_{ui} \, = \, h^{-1}\left( \argmax_{r \in \SSS'} \, \msum_{j \in \NN_u(i)}\!\!\! \delta(h(r_{uj}) = r) \, w_{ij} \right).
\eeq


\subsection{User-based VS item-based recommendation}\label{sec:ub-vs-ib-recommendation}

When choosing between the implementation of a user-based and an item-based neighborhood recommender system, five criteria should be considered:
\begin{itemize}
 
\item \textbf{Accuracy:} The accuracy of neighborhood recommendation methods depends mostly on the ratio between the number of users and items in the system. As will be presented in the Section \ref{sec:sim-weight-computation}, the similarity between two users in user-based methods, which determines the neighbors of a user, is normally obtained by comparing the ratings made by these users on the same items. Consider a system that has $10,000$ ratings made by $1,000$ users on $100$ items, and suppose, for the purpose of this analysis, that the ratings are distributed uniformly over the items\footnote{The distribution of ratings in real-life data is normally skewed, i.e. most ratings are given to a small proportion of items.}. Following Table \ref{table:user-vs-item-based-accuracy}, the average number of users available as potential neighbors is roughly $650$. However, the average number of common ratings used to compute the similarities is only $1$. On the other hand, an item-based method usually computes the similarity between two items by comparing ratings made by the same user on these items. Assuming once more a uniform distribution of ratings, we find an average number of potential neighbors of $99$ and an average number of ratings used to compute the similarities of $10$.

\itempar In general, a small number of high-confidence neighbors is by far preferable to a large number of neighbors for which the similarity weights are not trustable. In cases where the number of users is much greater than the number of items, such as large commercial systems like \emph{Amazon.com}, item-based methods can therefore produce more accurate recommendations \cite{fouss07,sarwar01}. Likewise, systems that have less users than items, e.g., a research paper recommender with thousands of users but hundreds of thousands of articles to recommend, may benefit more from user-based neighborhood methods \cite{herlocker04}.


\begin{table}[h!tb]
\begin{center}
\caption{The average number of neighbors and average number of ratings used in the computation of similarities for user-based and item-based neighborhood methods. A uniform distribution of ratings is assumed with average number of ratings per user $p = |\RR|/|\UU|$, and average number of ratings per item $q = |\RR|/|\II|$}
\label{table:user-vs-item-based-accuracy}
\begin{tabular}{x{2.0cm}||x{3.5cm}|x{2.5cm}}
\hline
           & Avg. neighbors & Avg. ratings \tn
\hline\hline
User-based & $(|\UU|-1)\left(1 - \left(\frac{|\II|-p}{|\II|}\right)^p\right)$ & $\frac{p^2}{|\II|}$ \tn
\hline
Item-based & $(|\II|-1)\left(1 - \left(\frac{|\UU|-q}{|\UU|}\right)^q\right)$ & $\frac{q^2}{|\UU|}$ \tn
\hline
\end{tabular}
\end{center}
\end{table}

\vspace{2mm}

\item \textbf{Efficiency:} As shown in Table \ref{table:user-vs-item-based-complexity}, the memory and computational efficiency of recommender systems also depends on the ratio between the number of users and items. Thus, when the number of users exceeds the number of items, as is it most often the case, item-based recommendation approaches require much less memory and time to compute the similarity weights (training phase) than user-based ones, making them more scalable. However, the time complexity of the online recommendation phase, which depends only on the number of available items and the maximum number of neighbors, is the same for user-based and item-based methods.

\itempar In practice, computing the similarity weights is much less expensive than the worst-case complexity reported in Table \ref{table:user-vs-item-based-complexity}, due to the fact that users rate only a few of the available items. Accordingly, only the non-zero similarity weights need to be stored, which is often much less than the number of user pairs. This number can be further reduced by storing for each user only the top $N$ weights, where $N$ is a parameter \cite{sarwar01} that is sufficient for satisfactory coverage on user-item pairs. In the same manner, the non-zero weights can be computed efficiently without having to test each pair of users or items, which makes neighborhood methods scalable to very large systems. 

\begin{table}[h!tb]
\begin{center}
\caption{The space and time complexity of user-based and item-based neighborhood methods, as a function of the maximum number of ratings per user $p = \max_{u} |\II_u|$, the maximum number of ratings per item $q = \max_{i} |\UU_i|$, and the maximum number of neighbors used in the rating predictions $k$.}
\label{table:user-vs-item-based-complexity}
\begin{tabular}{x{2.0cm}||x{1.5cm}|x{1.5cm}|x{1.5cm}}
\hline
           & \multirow{2}{*}{Space} & \multicolumn{2}{c}{Time} \tn           
           &                        & Training   & Online \tn
\hline\hline
User-based & $O(|\UU|^2)$ & $O(|\UU|^2 p)$ & $O(|\II| k)$ \tn 
Item-based & $O(|\II|^2)$ & $O(|\II|^2 q)$ & $O(|\II| k)$ \tn 
\hline
\end{tabular}
\end{center}
\end{table}

\vspace{2mm}

\item \textbf{Stability:} The choice between a user-based and an item-based approach also depends on the frequency and amount of change in the users and items of the system. If the list of available items is fairly static in comparison to the users of the system, an item-based method may be preferable since the item similarity weights could then be computed at infrequent time intervals while still being able to recommend items to new users. On the contrary, in applications where the list of available items is constantly changing, e.g., an online article recommender, user-based methods could prove to be more stable. 

\vspace{2mm}

\item \textbf{Justifiability:} An advantage of item-based methods is that they can easily be used to justify a recommendation. Hence, the list of neighbor items used in the prediction, as well as their similarity weights, can be presented to the user as an explanation of the recommendation. By modifying the list of neighbors and/or their weights, it then becomes possible for the user to participate interactively in the recommendation process. User-based methods, however, are less amenable to this process because the active user does not know the other users serving as neighbors in the recommendation. 

\vspace{2mm}

\item \textbf{Serendipity:} In item-based methods, the rating predicted for an item is based on the ratings given to similar items. Consequently, recommender systems using this approach will tend to recommend to a user items that are related to those usually appreciated by this user. For instance, in a movie recommendation application, movies having the same genre, actors or director as those highly rated by the user are likely to be recommended. While this may lead to safe recommendations, it does less to help the user discover different types of items that he might like as much. 

\itempar Because they work with user similarity, on the other hand, user-based approaches are more likely to make serendipitous recommendations. This is particularly true if the recommendation is made with a small number of nearest-neighbors. For example, a user $A$ that has watched only comedies may be very similar to a user $B$ only by the ratings made on such movies. However, if $B$ is fond of a movie in a different genre, this movie may be recommended to $A$ through his similarity with $B$.
\end{itemize}

\section{Components of neighborhood methods}\label{sec:components-nb-methods}

In the previous section, we have seen that deciding between a regression and a classification rating prediction method, as well as choosing between a user-based or item-based recommendation approach, can have a significant impact on the accuracy, efficiency and overall quality of the recommender system. In addition to these crucial attributes, three very important considerations in the implementation of a neighborhood-based recommender system are 1) the normalization of ratings, 2) the computation of the similarity weights, and 3) the selection of neighbors. This section reviews some of the most common approaches for these three components, describes the main advantages and disadvantages of using each one of them, and gives indications on how to implement them.

\subsection{Rating normalization}\label{sec:rating-normalization}

When it comes to assigning a rating to an item, each user has its own personal scale. Even if an explicit definition of each of the possible ratings is supplied (e.g., 1=``strongly disagree,'' 2=``disagree,'' 3=``neutral,'' etc.), some users might be reluctant to give high/low scores to items they liked/disliked. Two of the most popular rating normalization schemes that have been proposed to convert individual ratings to a more universal scale are \emph{mean-centering} and \emph{$Z$-score}.

\subsubsection{Mean-centering}

The idea of mean-centering \cite{breese98,resnick94} is to determine whether a rating is positive or negative by comparing it to the mean rating. In user-based recommendation, a raw rating $r_{ui}$ is transformation to a mean-centered one $h(r_{ui})$ by subtracting to $r_{ui}$ the average $\ol{r}_u$ of the ratings given by user $u$ to the items in $\II_u$:
\begin{displaymath}
  h(r_{ui}) \, = \, r_{ui} - \ol{r}_u.
\end{displaymath}
Using this approach the user-based prediction of a rating $r_{ui}$ is obtained as
\beq\label{eqn:user-based-norm-pred}
 \hat{r}_{ui} \, = \, \ol{r}_u \, + \, 
        \frac{\ssum_{v \in \NN_i(u)}\!\!\!w_{uv} \, (r_{vi} - \ol{r}_v)}
             {\ssum_{v \in \NN_i(u)}\!\!\!|w_{uv}|}.
\eeq
In the same way, the \emph{item}-mean-centered normalization of $r_{ui}$ is given by
\begin{displaymath}
  h(r_{ui}) \, = \, r_{ui} - \ol{r}_i,
\end{displaymath}
where $\ol{r}_i$ corresponds to the mean rating given to item $i$ by user in $\UU_i$. This normalization technique is most often used in item-based recommendation, where a rating $r_{ui}$ is predicted as:
\beq\label{eqn:item-based-norm-pred}
 \hat{r}_{ui} \, = \, \ol{r}_i \, + \, 
        \frac{\ssum_{j \in \NN_u(i)}\!\!\!w_{ij} \, (r_{uj} - \ol{r}_j)}
             {\ssum_{j \in \NN_u(i)}\!\!\!|w_{ij}|}.
\eeq
An interesting property of mean-centering is that one can see right-away if the appreciation of a user for an item is positive or negative by looking at the sign of the normalized rating. Moreover, the module of this rating gives the level at which the user likes or dislikes the item.

\begin{example}
As shown in Figure \ref{fig:mean-centered-ratings}, although Diane gave an average rating of 3 to the movies ``Titanic'' and ``Forrest Gump,'' the user-mean-centered ratings show that her appreciation of these movies is in fact negative. This is because her ratings are high on average, and so, an average rating correspond to a low degree of appreciation. Differences are also visible while comparing the two types of mean-centering. For instance, the item-mean-centered rating of the movie ``Titanic'' is neutral, instead of negative, due to the fact that much lower ratings were given to that movie. Likewise, Diane's appreciation for ``The Matrix'' and John's distaste for ``Forrest Gump'' are more pronounced in the item-mean-centered ratings.
\end{example}

\begin{figure}[h!tb]
\begin{small}
\begin{center}
\emph{User} mean-centering:\\
\vspace{2mm}
\begin{tabular}{x{1cm}||x{1cm}|x{1cm}|x{1cm}|x{1cm}|x{1cm}}
\hline
\multirow{2}{*}{} & The    & \multirow{2}{*}{Titanic} & Die  & Forrest & \multirow{2}{*}{Wall-E}\tn
                  & Matrix &                          & Hard & Gump    & \tn
\hline\hline
John  &  2.50 & -1.50 &       & -0.50 & -0.50 \tn
Lucy  & -2.60 &  1.40 & -1.60 &  1.40 &  1.40 \tn
Eric  & -1.50 &       & -0.50 &  1.50 &  0.50 \tn
Diane &  0.25 & -0.75 &  1.25 & -0.75 &       \tn
\hline
\end{tabular}

\vspace{5mm}

\emph{Item} mean-centering:\\
\vspace{2mm}
\begin{tabular}{x{1cm}||x{1cm}|x{1cm}|x{1cm}|x{1cm}|x{1cm}}
\hline
\multirow{2}{*}{} & The    & \multirow{2}{*}{Titanic} & Die  & Forrest & \multirow{2}{*}{Wall-E}\tn
                  & Matrix &                          & Hard & Gump    & \tn
\hline\hline
John  &  2.00 & -2.00 &       & -1.75 & -1.67 \tn
Lucy  & -2.00 &  2.00 & -1.33 &  1.25 &  1.33 \tn
Eric  & -1.00 &       & -0.33 &  1.25 &  0.33 \tn
Diane &  1.00 &  0.00 &  1.67 & -0.75 &       \tn
\hline
\end{tabular}
\end{center}
\caption{The \emph{user} and \emph{item} mean-centered ratings of Figure \ref{fig:toy-example}.}
\label{fig:mean-centered-ratings}
\end{small}
\end{figure}

\subsubsection{Z-score normalization}

Consider, two users $A$ and $B$ that both have an average rating of $3$. Moreover, suppose that the ratings of $A$ alternate between $1$ and $5$, while those of $B$ are always $3$. A rating of $5$ given to an item by $B$ is more exceptional than the same rating given by $A$, and, thus, reflects a greater appreciation for this item. While mean-centering removes the offsets caused by the different perceptions of an average rating, $Z$-score normalization \cite{herlocker99} also considers the spread in the individual rating scales. Once again, this is usually done differently in user-based than in item-based recommendation. In user-based methods, the normalization of a rating $r_{ui}$ divides the \emph{user}-mean-centered rating by the standard deviation $\sigma_u$ of the ratings given by user $u$:
\begin{displaymath}
  h(r_{ui}) \, = \, \frac{r_{ui} - \ol{r}_u}{\sigma_u}.
\end{displaymath}
A user-based prediction of rating $r_{ui}$ using this normalization approach would therefore be obtained as
\beq\label{eqn:user-based-zscore-pred}
 \hat{r}_{ui} \, = \, \ol{r}_u \, + \, 
        \sigma_u \, \frac{\ssum_{v \in \NN_i(u)}\!\!\!w_{uv} \, (r_{vi} - \ol{r}_v)/\sigma_v}
             {\ssum_{v \in \NN_i(u)}\!\!\!|w_{uv}|}.
\eeq
Likewise, the $z$-score normalization of $r_{ui}$ in item-based methods divides the \emph{item}-mean-centered rating by the standard deviation of ratings given to item $i$:
\begin{displaymath}
  h(r_{ui}) \, = \, \frac{r_{ui} - \ol{r}_i}{\sigma_i}.
\end{displaymath}
The item-based prediction of rating $r_{ui}$ would then be
\beq\label{eqn:item-based-zscore-pred}
 \hat{r}_{ui} \, = \, \ol{r}_i \, + \, 
        \sigma_i \,\frac{\ssum_{j \in \NN_u(i)}\!\!\!w_{ij} \, (r_{uj} - \ol{r}_j)/\sigma_j}
             {\ssum_{j \in \NN_u(i)}\!\!\!|w_{ij}|}.
\eeq

\subsubsection{Choosing a normalization scheme}

In some cases, rating normalization can have undesirable effects. For instance, imagine the case of a user that gave only the highest ratings to the items he has purchased. Mean-centering would consider this user as ``easy to please'' and any rating below this highest rating (whether it is a positive or negative rating) would be considered as negative. However, it is possible that this user is in fact ``hard to please'' and carefully selects only items that he will like for sure. Furthermore, normalizing on a few ratings can produce unexpected results. For example, if a user has entered a single rating or a few identical ratings, his rating standard deviation will be $0$, leading to undefined prediction values. Nevertheless, if the rating data is not overly sparse, normalizing ratings has been found to consistently improve the predictions \cite{herlocker99,howe08}.

Comparing mean-centering with $Z$-score, as mentioned, the second one has the additional benefit of considering the variance in the ratings of individual users or items. This is particularly useful if the rating scale has a wide range of discrete values or if it is continuous. On the other hand, because the ratings are divided and multiplied by possibly very different standard deviation values, $Z$-score can be more sensitive than mean-centering and, more often, predict ratings that are outside the rating scale. Lastly, while an initial investigation found mean-centering and $Z$-score to give comparable results \cite{herlocker99}, subsequent analysis showed $Z$-score to have more significant benefits \cite{howe08}.   

Finally, if rating normalization is not possible or does not improve the results, another possible approach to remove the problems caused by the rating scale variance is \emph{preference-based filtering}. The particularity of this approach is that it focuses on predicting the relative preferences of users instead of absolute rating values. Since, an item preferred to another one remains so regardless of the rating scale, predicting relative preferences removes the need to normalize the ratings. More information on this approach can be found in \cite{cohen98,freund98,jin03b,jin03a}.

\subsection{Similarity weight computation}\label{sec:sim-weight-computation}

The similarity weights play a double role in neighborhood-based recommendation methods: 1) they allow to select trusted neighbors whose ratings are used in the prediction, and 2) they provide the means to give more or less importance to these neighbors in the prediction. The computation of the similarity weights is one of the most critical aspects of building a neighborhood-based recommender system, as it can have a significant impact on both its accuracy and its performance. 

\subsubsection{Correlation-based similarity}

A measure of the similarity between two objects $a$ and $b$, often used in information retrieval, consists in representing these objects in the form of a vector $\xx_a$ and $\xx_b$ and computing the \emph{Cosine Vector} (CV) (or \emph{Vector Space}) similarity \cite{balabanovic97,billsus00,lang95} between these vectors:
\begin{displaymath}
  \cos(\xx_a, \xx_b) \, = \, \frac{\tr{\xx_a} \xx_b}{||\xx_a||\cdot||\xx_b||}.
\end{displaymath}
In the context of item recommendation, this measure can be employed to compute user similarities by considering a user $u$ as a vector $\xx_u \in \mathfrak{R}^{|I|}$, where $\xx_{ui} = r_{ui}$ if user $u$ has rated item $i$, and $0$ otherwise. The similarity between two users $u$ and $v$ would then be computed as
\beq
 CV(u,v) \, = \, \cos(\xx_u,\xx_v) \, = \, 
              \frac{\ssum_{i \in \II_{uv}} r_{ui} \, r_{vi}}
                    {\sqrt{\ssum_{i \in \II_u} r_{ui}^2 \ssum_{j \in \II_v} r_{vj}^2}},
\eeq
where $I_{uv}$ once more denotes the items rated by both $u$ and $v$. A problem with this measure is that is does not consider the differences in the mean and variance of the ratings made by users $u$ and $v$. 

A popular measure that compares ratings where the effects of mean and variance have been removed is the \emph{Pearson Correlation} (PC) similarity:
\beq\label{eqn:user-based-pc}
 \mathrm{PC}(u,v) \, = \, \frac{\ssum_{i \in \II_{uv}} (r_{ui} - \ol{r}_u) (r_{vi} - \ol{r}_v)}
                             {\sqrt{\ssum_{i \in \II_{uv}} (r_{ui} - \ol{r}_u)^2 
                                    \ssum_{i \in \II_{uv}} (r_{vi} - \ol{r}_v)^2}}.
\eeq
Note that this is different from computing the CV similarity on the $Z$-score normalized ratings, since the standard deviation of the ratings in evaluated only on the common items $I_{uv}$, not on the entire set of items rated by $u$ and $v$, i.e. $\II_u$ and $\II_v$. The same idea can be used to obtain similarities between two items $i$ and $j$ \cite{deshpande04,sarwar01}, this time by comparing the ratings made by users that have rated both these items:  
\beq\label{eqn:item-based-pc}
 \mathrm{PC}(i,j) \, = \, \frac{\ssum_{u \in \UU_{ij}} (r_{ui} - \ol{r}_i) (r_{uj} - \ol{r}_j)}
                             {\sqrt{\ssum_{u \in \UU_{ij}} (r_{ui} - \ol{r}_i)^2 
                                    \ssum_{u \in \UU_{ij}} (r_{uj} - \ol{r}_j)^2}}.
\eeq
While the sign of a similarity weight indicates whether the correlation is direct or inverse, its magnitude (ranging from $0$ to $1$) represents the strength of the correlation.

\begin{example}
The similarities between the pairs of users and items of our toy example, as computed using PC similarity, are shown in Figure \ref{fig:pc-sim-example}.  We can see that Lucy's taste in movies is very close to Eric's (similarity of $0.922$) but very different from John's (similarity of $-0.938$). This means that Eric's ratings can be trusted to predict Lucy's, and that Lucy should discard John's opinion on movies or consider the opposite. We also find that the people that like ``The Matrix'' also like ``Die Hard'' but hate ``Wall-E''. Note that these relations were discovered without having any knowledge of the genre, director or actors of these movies.
\end{example}

\begin{figure}[h!tb]
\begin{small}
\begin{center}
\emph{User-based} Pearson correlation
\vspace{2mm}

\begin{tabular}{x{1cm}||x{1cm}|x{1cm}|x{1cm}|x{1cm}}
\hline
      &  John  &  Lucy  &  Eric  & Diane \tn
\hline\hline  
John  &  1.000 & -0.938 & -0.839 &  0.659 \tn
Lucy  & -0.938 &  1.000 &  0.922 & -0.787 \tn
Eric  & -0.839 &  0.922 &  1.000 & -0.659 \tn
Diane &  0.659 & -0.787 & -0.659 &  1.000 \tn
\hline
\end{tabular}

\vspace{5mm}
\emph{Item-based} Pearson correlation
\vspace{2mm}

\begin{tabular}{x{1.75cm}||x{1cm}|x{1cm}|x{1cm}|x{1cm}|x{1cm}}
\hline
\multirow{2}{*}{} & The    & \multirow{2}{*}{Titanic} & Die  & Forrest & \multirow{2}{*}{Wall-E}\tn
                  & Matrix &                          & Hard & Gump    & \tn
  \hline\hline 
Matrix       &  1.000 & -0.943 &  0.882 & -0.974 & -0.977 \tn
Titanic      & -0.943 &  1.000 & -0.625 &  0.931 &  0.994 \tn
Die Hard     &  0.882 & -0.625 &  1.000 & -0.804 & -1.000 \tn
Forrest Gump & -0.974 &  0.931 & -0.804 &  1.000 &  0.930 \tn
Wall-E       & -0.977 &  0.994 & -1.000 &  0.930 &  1.000 \tn
\hline
\end{tabular}
\caption{The \emph{user} and \emph{item} PC similarity for the ratings of Figure \ref{fig:toy-example}.}
\label{fig:pc-sim-example}
\end{center}
\end{small}
\end{figure}

The differences in the rating scales of individual users are often more pronounced than the differences in ratings given to individual items. Therefore, while computing the item similarities, it may be more appropriate to compare ratings that are centered on their \emph{user} mean, instead of their \emph{item} mean. The \emph{Adjusted Cosine} (AC) similarity \cite{sarwar01}, is a modification of the PC item similarity which compares user-mean-centered ratings:
\begin{displaymath}
  AC(i,j) \, = \, \frac{\ssum_{u \in \UU_{ij}} (r_{ui} - \ol{r}_u)(r_{uj} - \ol{r}_u)}
             {\sqrt{\ssum_{u \in \UU_{ij}} (r_{ui} - \ol{r}_u)^2 \ssum_{u \in \UU_{ij}}(r_{uj} - \ol{r}_u)^2}}.
\end{displaymath}
In some cases, AC similarity has been found to outperform PC similarity on the prediction of ratings using an item-based method \cite{sarwar01}.

\subsubsection{Other similarity measures}

Several other measures have been proposed to compute similarities between users or items. One of them is the \emph{Mean Squared Difference} (MSD) \cite{shardanand95}, which evaluate the similarity between two users $u$ and $v$ as the inverse of the average squared difference between the ratings given by $u$ and $v$ on the same items:
\beq\label{eqn:user-based-msd}
 \mathrm{MSD}(u,v) \, = \, \frac{|\II_{uv}|}{\ssum_{i \in \II_{uv}} (r_{ui} - r_{vi})^2}.
\eeq
While it could be modified to compute the differences on normalized ratings, the MSD similarity is limited compared to PC similarity because it does not allows to capture negative correlations between user preferences or the appreciation of different items. Having such negative correlations may improve the rating prediction accuracy \cite{herlocker02}.

Another well-known similarity measure is the \emph{Spearman Rank Correlation} (SRC) \cite{kendall90rank}. While PC uses the rating values directly, SRC instead considers the ranking of these ratings. Denote by $k_{ui}$ the rating rank of item $i$ in user $u$'s list of rated items (tied ratings get the average rank of their spot). The SRC similarity between two users $u$ and $v$ is evaluated as:
\beq\label{eqn:user-based-src}
 \mathrm{SRC}(u,v) \, = \, \frac{\ssum_{i \in \II_{uv}} (k_{ui} - \ol{k}_u) (k_{vi} - \ol{k}_v)}
                             {\sqrt{\ssum_{i \in \II_{uv}} (k_{ui} - \ol{k}_u)^2 
                                    \ssum_{i \in \II_{uv}} (k_{vi} - \ol{k}_v)^2}},
\eeq
where $\ol{k}_u$ is the average rank of items rated by $u$.

The principal advantage of SRC is that it avoids the problem of rating normalization, described in the last section, by using rankings. On the other hand, this measure may not be the best one when the rating range has only a few possible values, since that would create a large number of tied ratings. Moreover, this measure is typically more expensive than PC as ratings need to be sorted in order to compute their rank.  

Table \ref{fig:sim-mae-comparison} shows the user-based prediction accuracy (MAE) obtained with MSD, SRC and PC similarity measures, on the \emph{MovieLens}\footnote{\url{http://www.grouplens.org/}} dataset \cite{herlocker02}. Results are given for different values of $k$, which represents the maximum number of neighbors used in the predictions. For this data, we notice that MSD leads to the least accurate predictions, possibly due to the fact that it does not take into account negative correlations. Also, these results show PC to be slightly more accurate than SRC. Finally, although PC has been generally recognized as the best similarity measure, see e.g. \cite{herlocker02}, subsequent investigation has shown that the performance of such measure depended greatly on the data \cite{howe08}.

\begin{table}[h!tb]
\begin{center}
\caption{The rating prediction accuracy (MAE) obtained on the \emph{MovieLens} dataset using the Mean Squared Difference (MSD), Spearman Rank Correlation and Pearson Correaltion (PC) similarity measures. Results are shown for predictions using an increasing number of neighbors $k$.}
\label{fig:sim-mae-comparison}
\begin{small}
\begin{tabular}{x{1cm}||x{1cm}|x{1cm}|x{1cm}}
\hline
$k$ & MSD & SRC & PC \tn
\hline\hline
5   & 0.7898 & 0.7855 & 0.7829 \tn
10  & 0.7718 & 0.7636 & 0.7618 \tn
20  & 0.7634 & 0.7558 & 0.7545 \tn
60  & 0.7602 & 0.7529 & 0.7518 \tn
80  & 0.7605 & 0.7531 & 0.7523 \tn
100 & 0.7610 & 0.7533 & 0.7528 \tn
\hline
\end{tabular}
\end{small}
\end{center}
\end{table}

\subsubsection{Considering the significance of weights}\label{sec:accounting-for-significance}

Because the rating data is frequently sparse in comparison to the number of users and items of a system, it is often the case that similarity weights are computed using only a few ratings given to common items or made by the same users. For example, if the system has $10,000$ ratings made by $1,000$ users on $100$ items (assuming a uniform distribution of ratings), Table \ref{table:user-vs-item-based-accuracy} shows us that the similarity between two users is computed, on average, by comparing the ratings given by these users to a \emph{single} item. If these few ratings are equal, then the users will be considered as ``fully similar'' and will likely play an important role in each other's recommendations. However, if the users' preferences are in fact different, this may lead to poor recommendations.

Several strategies have been proposed to take into account the \emph{significance} of a similarity weight. The principle of these strategies is essentially the same: reduce the magnitude of a similarity weight when this weight is computed using only a few ratings. For instance, in \emph{Significance Weighting} \cite{herlocker99,ma07}, a user similarity weight $w_{uv}$ is penalized by a factor proportional to the number of commonly rated item, if this number is less than a given parameter $\gamma > 0$:
\beq
  w'_{uv} \, = \, \frac{\min\{|\II_{uv}|, \, \gamma\}}{\gamma} \times w_{uv}.
\eeq
Likewise, an item similarity $w_{ij}$, obtained from a few ratings, can be adjusted as
\beq
  w'_{ij} \, = \, \frac{\min\{|\UU_{ij}|, \, \gamma\}}{\gamma} \times w_{ij}.
\eeq
In \cite{herlocker99,herlocker02}, it was found that using $\gamma \geq 25$ could significantly improve the accuracy of the predicted ratings, and that a value of $50$ for $\gamma$ gave the best results. However, the optimal value for this parameter is data dependent and should be determined using a cross-validation approach.

A characteristic of significance weighting is its use of a threshold $\gamma$ determining when a weight should be adjusted. A more continuous approach, described in \cite{bell07a}, is based on the concept of \emph{shrinkage} where a weak or biased estimator can be improved if it is ``shrunk'' toward a null-value. This approach can be justified using a Bayesian perspective, where the best estimator of a parameter is the posterior mean, corresponding to a linear combination of the prior mean of the parameter (null-value) and an empirical estimator based fully on the data. In this case, the parameters to estimate are the similarity weights and the null value is zero. Thus, a user similarity $w_{uv}$ estimated on a few ratings is shrunk as
\beq
  w'_{uv} \, = \, \frac{|\II_{uv}|}{|\II_{uv}| + \beta} \times w_{uv},
\eeq
where $\beta > 0$ is a parameter whose value should also be selected using cross-validation. In this approach, $w_{uv}$ is shrunk proportionally to $\beta/|I_{uv}|$, such that almost no adjustment is made when $|\II_{uv}| \gg \beta$. Item similarities can be shrunk in the same way:
\beq
  w'_{ij} \, = \, \frac{|\UU_{ij}|}{|\UU_{ij}| + \beta} \times w_{ij},
\eeq
As reported in \cite{bell07a}, a typical value for $\beta$ is 100.

\subsubsection{Considering the variance of ratings}

Ratings made by two users on universally liked/disliked items may not be as informative as those made for items with a greater rating variance. For instance, most people like classic movies such as ``The Godfather'' so basing the weight computation on such movies would produce artificially high values. Likewise, a user that always rates items in the same way may provide less predictive information than one whose preferences vary from one item to another. 

A recommendation approach that addresses this problem is the \emph{Inverse User Frequency} \cite{breese98}. Based on the information retrieval notion of \emph{Inverse Document Frequency} (IDF), a weight $\lambda_i$ is given to each item $i$, in proportion to the log-ratio of users that have rated $i$:
\begin{displaymath}
  \lambda_i \, = \, \log\frac{|\UU|}{|\UU_i|}.
\end{displaymath}
In the \emph{Frequency-Weighted Pearson Correlation} (FWPC), the correlation between the ratings given by two users $u$ and $v$ to an item $i$ is weighted by $\lambda_i$:
\beq\label{eqn:fw-user-based-pc}
 \mathrm{FWPC}(u,v) \, = \, \frac{\ssum_{i \in \II_{uv}} \lambda_i (r_{ui} - \ol{r}_u) (r_{vi} - \ol{r}_v)}
                             {\sqrt{\ssum_{i \in \II_{uv}} \lambda_i (r_{ui} - \ol{r}_u)^2 
                                    \ssum_{i \in \II_{uv}} \lambda_i (r_{vi} - \ol{r}_v)^2}}.
\eeq
This approach, which was found to improve the prediction accuracy of a user-based recommendation method \cite{breese98}, could also be adapted to the computation of item similarities. More advanced strategies have also been proposed to consider rating variance. One of these strategies, described in \cite{jin04}, computes the factors $\lambda_i$ by maximizing the average similarity between users. 

\subsubsection{Considering the target item}

If the goal is to predict ratings with a user-based method, more reliable correlation values can be obtained if the target item is considered in their computation. In \cite{baltrunas2009item}, the user-based PC similarity is extended by weighting the summation terms corresponding to an item $i$ by the similarity between $i$ and the target item $j$: 
\beq\label{eqn:WPCCnorm}
\mathrm{WPC}_j(u,v) \, = \, \frac{\ssum_{i \in \II_{uv}} w_{ij} \, (r_{ui} - \ol{r}_u) (r_{vi} - \ol{r}_v)}
                             {\sqrt{\ssum_{i \in \II_{uv}} w_{ij} \, (r_{ui} - \ol{r}_u)^2 
                                    \ssum_{i \in \II_{uv}} w_{ij} \, (r_{vi} - \ol{r}_v)^2}}.
\eeq
The item weights $w_{ij}$ can be computed using PC similarity or obtained by considering the items' content (e.g., the common genres for movies). Other variations of this similarity metric and their impact on the prediction accuracy are described in \cite{baltrunas2009item}. Note, however, that this model may require to recompute the similarity weights for each predicted rating, making it less suitable for online recommender systems.

\subsection{Neighborhood selection}

The number of nearest-neighbors to select and the criteria used for this selection can also have a serious impact on the quality of the recommender system. The selection of the neighbors used in the recommendation of items is normally done in two steps: 1) a global filtering step where only the most likely candidates are kept, and 2) a per prediction step which chooses the best candidates for this prediction. 

\subsubsection{Pre-filtering of neighbors}

In large recommender systems that can have millions of users and items, it is usually not possible to store the (non-zero) similarities between each pair of users or items, due to memory limitations. Moreover, doing so would be extremely wasteful as only the most significant of these values are used in the predictions. The pre-filtering of neighbors is an essential step that makes neighborhood-based approaches practicable by reducing the amount of similarity weights to store, and limiting the number of candidate neighbors to consider in the predictions. There are several ways in which this can be accomplished:

\begin{itemize}
\item \textbf{Top-$N$ filtering:} For each user or item, only a list of the $N$ nearest-neighbors and their respective similarity weight is kept. To avoid problems with efficiency or accuracy, $N$ should be chosen carefully. Thus, if $N$ is too large, an excessive amount of memory will be required to store the neighborhood lists and predicting ratings will be slow. On the other hand, selecting a too small value for $N$ may reduce the coverage of the recommendation method, which causes some items to be never recommended.

\vspace{2mm}

\item \textbf{Threshold filtering:} Instead of keeping a fixed number of nearest-neighbors, this approach keeps all the neighbors whose similarity weight's magnitude is greater than a given threshold $w_\text{min}$. While this is more flexible than the previous filtering technique, as only the most significant neighbors are kept, the right value of $w_\text{min}$ may be difficult to determine.

\vspace{2mm}

\item \textbf{Negative filtering:} In general, negative rating correlations are less reliable than positive ones. Intuitively, this is because strong positive correlation between two users is a good indicator of their belonging to a common group (e.g., teenagers, science-fiction fans, etc.). However, although negative correlation may indicate membership to different groups, it does not tell how different are these groups, or whether these groups are compatible for some other categories of items. While certain  experimental investigations \cite{herlocker99,herlocker04} have found negative correlations to provide no significant improvement in the prediction accuracy, in certain settings they seem to have a positive effect (see e.g., \cite{EASE}). Whether such correlations can be discarded depends on the data and should be examined on a case-by-case basis. 
\end{itemize}

Note that these three filtering approaches are not exclusive and can be combined to fit the needs of the recommender system. For instance, one could discard all negative similarities \emph{as well as} those that are not in the top-$N$ lists.  

\subsubsection{Neighbors in the predictions}

Once a list of candidate neighbors has been computed for each user or item, the prediction of new ratings is normally made with the $k$-nearest-neighbors, that is, the $k$ neighbors whose similarity weight has the greatest magnitude. The choice of $k$ can also have a significant impact on the accuracy and performance of the system.

As shown in Table \ref{fig:sim-mae-comparison}, the prediction accuracy observed for increasing values of $k$ typically follows a \emph{concave} function. Thus, when the number of neighbors is restricted by using a small $k$ (e.g., $k < 20$), the prediction accuracy is normally low. As $k$ increases, more neighbors contribute to the prediction and the variance introduced by individual neighbors is averaged out. As a result, the prediction accuracy improves. Finally, the accuracy usually drops when too many neighbors are used in the prediction (e.g., $k > 50$), due to the fact that the few strong local relations are ``diluted'' by the many weak ones. Although a number of neighbors between $20$ to $50$ is most often described in the literature, see e.g. \cite{herlocker02,herlocker04}, the optimal value of $k$ should be determined by cross-validation.

On a final note, more serendipitous recommendations may be obtained at the cost of a decrease in accuracy, by basing these recommendations on a few very similar users. For example, the system could find the user most similar to the active one and recommend the new item that has received the highest rated from this user.

\section{Advanced techniques}\label{sec:advanced-techniques}

The neighborhood approaches based on rating correlation, such as the ones presented in the previous sections, have three important limitations:
\begin{itemize}
\item \textbf{Limited Expressiveness:} Traditional neighborhood-based methods  determine the neighborhood  of  users or items using  some predefined  similarity measure like cosine or PC. Recommendation algorithms that rely on such similarity measures have been shown to enjoy remarkable recommendation accuracy in certain settings. However their performance can vary considerably depending on whether the chosen similarity measures conform with the latent characteristics of the dataset onto which they are applied. 

\vspace{2mm}

\item \textbf{Limited coverage:} Because rating correlation measures the similarity between two users by comparing their ratings for the same items, users can be neighbors \emph{only if} they have rated common items. This assumption is very limiting, as users having rated a few or no common items may still have similar preferences. Moreover, since only items rated by neighbors can be recommended, the coverage of such methods can also be limited. This limitation also applies when two items have only a few or no co-ratings. 

\vspace{2mm}

\item \textbf{Sensitivity to sparse data:} Another consequence of rating correlation, addressed briefly in Section \ref{sec:ub-vs-ib-recommendation}, is the fact that the accuracy of neighborhood-based recommendation methods suffers from the lack of available ratings. Sparsity is a problem common to most recommender systems due to the fact that users typically rate only a small proportion of the available items \cite{billsus98,good99,sarwar98,sarwar00b}. This is aggravated by the fact that users or items newly added to the system may have no ratings at all, a problem known as \emph{cold-start} \cite{schein02}. When the rating data is sparse, two users or items are unlikely to have common ratings, and consequently, neighborhood-based approaches will predict ratings using a very limited number of neighbors. Moreover, similarity weights may be computed using only a small number of ratings, resulting in biased recommendations (see Section \ref{sec:accounting-for-significance} for this problem).
\end{itemize}

A common solution for  {latter} problems is to fill the missing ratings with default values \cite{breese98,deshpande04}, such as the middle value of the rating range, or the average user or item rating. A more reliable approach is to use content information to fill out the missing ratings \cite{degemmis07,good99,konstan97,melville02}. For instance, the missing ratings can be provided by autonomous agents called \emph{filterbots} \cite{good99,konstan97}, that act as ordinary users of the system and rate items based on some specific characteristics of their content. The missing ratings can instead be predicted by a content-based approach \cite{melville02}. Furthermore, content similarity can also be used ``instead of'' or ``in addition to'' rating correlation similarity to find the nearest-neighbors employed in the predictions \cite{balabanovic97,li04,pazzani99,soboroff99}. Finally, data sparsity can also be tackled by acquiring new ratings with active learning techniques. In such techniques, the system interactively queries the user to gain a better understanding of his or her preferences. A more detailed presentation of interactive and session-based techniques is given in Chapter~\ref{4-session-rs} of this book. These solutions, however, also have their own drawbacks. For instance, giving a default values to missing ratings may induce bias in the recommendations. Also, item content may not be available to compute ratings or similarities. 

This section presents two approaches that aim to tackle the  {aforementioned challenges}:  \emph{learning-based} and \emph{graph-based}  methods.

\subsection{Learning-based methods}

In the methods of this family the similarity or affinity between users and items is obtained by defining a parametric model that describes the relation between users, items or both, and then fits the model parameters through an optimization procedure. 

Using a learning-based method has significant advantages. First, such methods can capture high-level patterns  and trends in the data, are generally more robust to outliers, and are known to generalize better than approaches solely based on local relations. In recommender systems, this translates into greater accuracy and stability in the recommendations \cite{koren08}. Also, because the relations between users and items are encoded in a limited set of parameters, such methods normally require less memory than other types of approaches. Finally, since the parameters are usually learned offline, the online recommendation process is generally faster.

Learning-based methods that use neighborhood or similarity information can be divided in two categories: factorization methods and adaptive neighborhood learning methods. These categories are presented in the following sections.

\subsubsection{Factorization methods}

Factorization methods \cite{bell07a,billsus98, puresvd, goldberg01,koren08, eigenrec, sarwar00b,takacs08,takacs09} address the problems of limited coverage and sparsity by projecting users and items into a reduced latent space that captures their most salient features. Because users and items are compared in this dense subspace of high-level features, instead of the ``rating space,'' more meaningful relations can be discovered. In particular, a relation between two users can be found, even though these users have rated different items. As a result, such methods are generally less sensitive to sparse data \cite{bell07a,billsus98,sarwar00b}.

There are essentially two ways in which factorization can be used to improve recommender systems: 1) factorization of a sparse \emph{similarity} matrix, and 2) factorization of a user-item \emph{rating} matrix.

\paragraph{\textbf{Factorizing the similarity matrix}}

Neighborhood similarity measures like the correlation similarity are usually very sparse since the average number of ratings per user is much less than the total number of items. A simple solution to densify a sparse similarity matrix is to compute a low-rank approximation of this matrix with a factorization method. 

Let $W$ be a symmetric matrix of rank $n$ representing either user or item similarities. To simplify the presentation, we will suppose the latter case. We wish to approximate $W$ with a matrix $\hat{W} = Q\tr{Q}$ of lower rank $k < n$, by minimizing the following objective:
\begin{eqnarray*}
    E(Q) & \, = \, & ||W - Q \tr{Q}||^2_F \\
                 & \, = \, & \msum_{i,j} \left(w_{ij} - \qq_i\tr{\qq_j}\right)^2,
\end{eqnarray*}
where $||M||_F = \sqrt{\sum_{i,j} m^2_{ij}}$ is the matrix Frobenius norm. Matrix $\hat{W}$ can be seen as a ``compressed'' and less sparse version of $W$. Finding the factor matrix $Q$ is equivalent to computing the eigenvalue decomposition of $W$:
\begin{displaymath}
  W \, = \, V D \tr{V},
\end{displaymath}
where $D$ is a diagonal matrix containing the $|\II|$ eigenvalues of $W$, and $V$ is a $|\II|\!\times\!|\II|$ orthogonal matrix containing the corresponding eigenvectors. Let $V_k$ be a matrix formed by the $k$ principal (normalized) eigenvectors of $W$, which correspond to the axes of the $k$-dimensional latent subspace. The coordinates $\qq_i \in \mathfrak{R}^k$ of an item $i$ in this subspace is given by the $i$-th row of matrix $Q = V_k D_k^{1/2}$. Furthermore, the item similarities computed in this latent subspace are given by matrix 
\begin{eqnarray}
 \hat{W} & \, = \, & Q \tr{Q}\nonumber \\
    & \, = \, & V_k D_k \tr{V}_k. 
\end{eqnarray}

This approach was used to recommend jokes in the Eigentaste system \cite{goldberg01}. In Eigentaste, a matrix $W$ containing the PC similarities between pairs of items is decomposed to obtain the latent subspace defined by the $k$ principal eigenvectors of $W$. A user $u$, represented by the $u$-th row $\rr_u$ of the rating matrix $R$, is projected in the plane defined by $V_k$:
\begin{displaymath}
  \rr'_u \, = \, \rr_u V_k. 
\end{displaymath}
In an offline step, the users of the system are clustered in this subspace using a recursive subdivision technique. Then, the rating of user $u$ for an item $i$ is evaluated as the mean rating for $i$ made by users in the same cluster as $u$. This strategy is related to the well-known spectral clustering method \cite{shi2000normalized}. 

\paragraph{\textbf{Factorizing the rating matrix}}

The problems of cold-start and limited coverage can also be alleviated by factorizing the user-item rating matrix. Once more, we want to approximate the $|\UU|\!\times\!|\II|$ rating matrix $R$ of rank $n$ by a matrix $\hat{R} = P\tr{Q}$ of rank $k < n$, where $P$ is a $|\UU|\!\times\!k$ matrix of \emph{users} factors and $Q$ a $|\II|\!\times\!k$ matrix of \emph{item} factors. This task can be formulated as finding matrices $P$ and $Q$ which minimize the following function:
\begin{eqnarray*}
  E(P,Q) & \, = \, & ||R - P \tr{Q}||^2_F \\
                 & \, = \, & \msum_{u,i} \left(r_{ui} - \pp_u\tr{\qq_i}\right)^2.
\end{eqnarray*}
The optimal solution can be obtained by the Singular Value Decomposition (SVD) of $R$: $P = U_k D_k^{1/2}$ and $Q = V_k D_k^{1/2}$, where $D_k$ is a diagonal matrix containing the $k$ largest singular values of $R$, and $U_k, V_k$ respectively contain the left and right singular vectors corresponding to these values.

However, there is significant problem with applying SVD directly to the rating matrix $R$: most values $r_{ui}$ of $R$ are undefined, since there may not be a rating given to $i$ by $u$. Although it is possible to assign a default value to $r_{ui}$, as mentioned above, this would introduce a bias in the data. More importantly, this would make the large matrix $R$ dense and, consequently, render impractical the SVD decomposition of $R$. A common solution to this problem is to learn the model parameters using only the known ratings \cite{bell07a,koren08,takacs07,takacs09}. For instance, suppose the rating of user $u$ for item $i$ is estimated as
\beq\label{eqn:svd-rating}
  \hat{r}_{ui} \, = \, b_u \, + \, b_i \, + \, \pp_u \tr{\qq}_i,
\eeq
where $b_u$ and $b_i$ are parameters representing the user and item rating biases. The model paremeters can be learned by minimizing the following objective function:
\beq\label{eqn:svd-factor}
  E(P,Q,\bb) \, = \, \msum_{r_{ui} \in \RR} (r_{ui} - \hat{r}_{ui})^2  
                            \, + \, \lambda\left(||\pp_u||^2 +  ||\qq_i||^2 + b^2_u + b^2_i\right).
\eeq
The second term of the function is as a regularization term added to avoid overfitting. Parameter $\lambda$ controls the level of regularization. A more comprehensive description of this recommendation approach can be found in Chapter~\ref{15-collab-filt-rs} of this book. 

The SVD model of Equation \ref{eqn:svd-rating} can be transformed into a similarity-based method by supposing that the profile of a user $u$ is determined implicitly by the items he or she has rated. Thus, the factor vector of $u$ can be defined as a weighted combination of the factor vectors $\sss_j$ corresponding to the items $j$ rated by this user:
\beq\label{eqn:user-factor-est}
  \pp_u \, = \, |\II_u|^{-\alpha} \ssum_{j \in \II_u} c_{uj} \, \sss_j.
\eeq
In this formulation, $\alpha$ is a normalization constant typically set to $\alpha=1/2$, and $c_{uj}$ is a weight representing the contribution of item $j$ to the profile of $u$. For instance, in the SVD++ model \cite{koren08} this weight is defined as the bias corrected rating of $u$ for item $j$: $c_{uj} = r_{ui} - b_u - b_j$. Other approaches, such as the FISM 
\cite{Kabbur2013} and NSVD \cite{paterek07} models, instead use constant weights: $c_{uj} = 1$. 

Using the formulation of Equation \ref{eqn:user-factor-est}, a rating $r_{ui}$ is predicted as
\beq
  \hat{r}_{ui} \, = \, b_u \, + \, b_i \, + \, |\II_u|^{-\alpha} \ssum_{j \in \II_u} c_{uj} \, \sss_j \tr{\qq}_i.
\eeq
Like the standard SVD model, the parameters of this model can be learned by minimizing the objective function of Equation (\ref{eqn:svd-factor}), for instance, using gradient descent optimization. 

Note that, instead of having both user and item factors, we now have two different sets of item factors, i.e., $\qq_i$ and $\sss_j$. These vectors can be interpreted as the factors of an asymmetric item-item similarity matrix $W$, where
\beq
   w_{ij} \, = \, \sss_i \tr{\qq}_j.
\eeq
As mentioned in \cite{koren08}, this similarity-based factorization approach has several advantages over the traditional SVD model. First, since there are typically more users than items in a recommender system, replacing the user factors by a combination of item factors reduces the number of parameters in the model, which makes the learning process faster and more robust. Also, by using item similarities instead of user factors, the system can handle new users without having to re-train the model. Finally, as in item-similarity neighborhood methods, this model makes it possible to justify a rating to a user by showing this user the items that were most involved in the prediction.

In FISM \cite{Kabbur2013}, the prediction of a rating $r_{ui}$ is made without considering the factors of $i$:
\beq
  \hat{r}_{ui} \, = \, b_u \, + \, b_i \, + \, 
  		\big(|\II_u| - 1\big)^{-\alpha} \ssum_{j \in \II_u \! \setminus \! \{i\}} \sss_j \tr{\qq}_i.
\eeq
This modification, which corresponds to ignoring the diagonal entries in the item similarity matrix, avoids the problem of having an item recommending itself and has been shown to give better performance when the number of factors is high.

\subsubsection{Neighborhood-learning methods}

Standard neighborhood-based recommendation algorithms determine the neighborhood of users or items directly from the data, using some pre-defined similarity measure like PC.
However, subsequent developments in the field of item recommendation have shown the advantage of learning the neighborhood automatically from the data, instead of using a pre-defined similarity measure \cite{Koenigstein2013,koren08,Natarajan2013,Rendle2009}.

\paragraph{\textbf{Sparse linear neighborhood model}}

A representative neighborhood-learning recommendation method is the \SLIM algorithm, developed by Ning \etal~\cite{Ning2011}. 
In \SLIM, a new rating is predicted as a sparse aggregation of existing ratings in a user's profile, 
\beq\label{eqn:slimpred}
  \hat{r}_{ui} \, = \, \rr_u \tr{\ww}_i, 
\eeq
where $\rr_u$ is the $u$-th row of the rating matrix $R$ and $\ww_j$ is a sparse row vector containing $|\II|$ aggregation 
coefficients. Essentially, the non-zero entries in $\ww_i$ correspond to the neighbor items of an item $i$. 

The neighborhood parameters are learned by minimizing the squared prediction error. Standard regularization and sparsity are enforced by penalizing the $\ell_2$-norm and $\ell_1$-norm of the parameters. The combination of these two types of regularizers in a regression problem is known as elastic net regularization \cite{Zou2005}. This learning process can be expressed as the following optimization problem:
\beqa \label{eqn:slim-opt}
  \displaystyle{
    \begin{aligned}
      & \underset{W}{\text{minimize}}
      & & \frac{1}{2}\| R - R W \|^2_F 
         + \frac{\beta}{2} \| W \|^2_F
         + \lambda \| W \|_1 \\
      & \text{subject to}
      & &  W \ge 0 \\
      & 
      & & \mbox{diag}(W) = 0. 
    \end{aligned}
  }
\eeqa
The constraint $\mbox{diag}(W) = 0$ is added to the model to avoid trivial solutions (e.g., $W$ corresponding to the identity matrix) and ensure that $r_{ui}$ is not used to compute $\hat{r}_{ui}$ during the recommendation process. Parameters $\beta$ and $\lambda$ control the amount of each type of regularization. Moreover, the non-negativity constraint on $W$ imposes the relations between neighbor items to be positive. Dropping the non-negativity as well as the sparsity constraints has been recently explored in~\cite{EASE}, and was shown to work well on several datasets with small number of items with respect to users. Note, however, that without the sparsity constraint the resulting model will be fully dense; a fact that imposes practical limitations on the applicability of such approaches in large item-space regimes.

\begin{figure}
    \centering
    \includegraphics[width=0.9\textwidth]{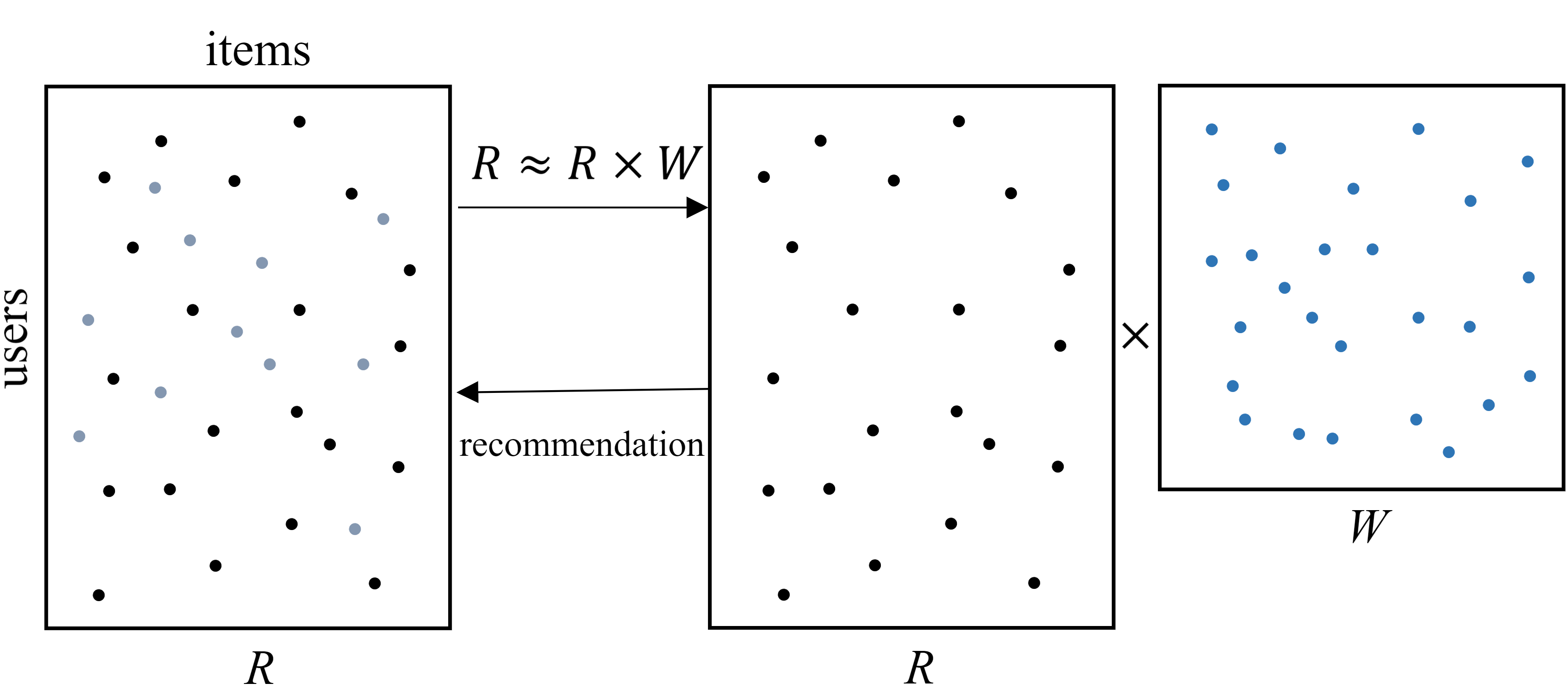}
    \caption{A simple illustration of \SLIM. The method works by first building an item-to-item model, based on $R$. Intuitively, this item model expresses each item (i.e., each column of the original rating matrix $R$) as a \textit{sparse linear combination} of the rest of the items (i.e., the other columns of $R$). Then, given $W$, new recommendations for a target user $u$ can be readily produced by multiplying the row  corresponding to user $u$ (i.e. the $u$-th row of $R$), with the learned item model, $W$.}
    \label{fig:slim}
\end{figure}

\paragraph{\textbf{Sparse neighborhood with side information}}

Side information, such as user profile attributes (e.g., age, gender, location) or item descriptions/tags, is becoming increasingly available in e-commerce applications. Properly exploited, this rich source of information can significantly improve the performance of conventional recommender systems \cite{Adams2010,Agarwal2011,Yoo2009,Singh2008}.

Item side information can be integrated in the \SLIM model by supposing that the co-rating profile of two items is correlated to the properties encoded in their side information \cite{r14}. To enforce such correlations in the model, an additional requirement is added, where both the user-item rating matrix $R$ and the item side information matrix $F$ should be reproduced by the same sparse linear aggregation. That is, in addition to satisfying $R \sim R W$, the coefficient matrix $W$ should also satisfy $F \sim F W$. This is achieved by solving the following optimization problem: 
\beqa
  \label{opt:cslim}
  \displaystyle{
    \begin{aligned}
      &   \underset{W}{\text{minimize}}
      &   & \frac{1}{2}\| R - RW \|^2_F \, + \, \frac{\alpha}{2}\| F - FW \|^2_F 
      	\, + \, \frac{\beta}{2} \| W \|^2_F \, + \, \lambda \| W \|_1 \; \\    
      & \text{subject to}
      &   &  W \ge 0, \;\\
      &
      &   & \mbox{diag}(W) = 0. \\
    \end{aligned}
  }
\eeqa
The parameter $\alpha$ is used to control the relative importance of the user-item rating information $R$ and the item side information $F$ when they are used to learn $W$. 

In some cases, requiring that the aggregation coefficients be the same for both $R$ and $F$ can be too strict. An alternate model relaxes this constraints by imposing these two sets of aggregation coefficients to be similar. Specifically, it uses an aggregation coefficient matrix $Q$ such that $F \sim F Q$ and $W \sim Q$. Matrices $W$ and $Q$ are learned as 
the minimizers of the following optimization problem:
%
\beqa
  \label{opt:rcslim}
  \displaystyle{
    \begin{aligned}
      &   \underset{W, Q}{\text{minimize}}
      &   & \frac{1}{2}\| R - RW \|^2_F  \, + \, \frac{\alpha}{2}\| F - FQ \|^2_F 
       \, + \, \frac{\beta_1}{2} \| W - Q\|^2_F \; \\
      &
      & & \quad \, + \, \frac{\beta_2}{2} \big(\| W \|^2_F + \| Q \|^2_F\big)       
       \, + \, \lambda \big( \| W \|_1 + \| Q \|_1\big) \;\\
      & \text{subject to}
      &   &  W, Q \ge 0, \\
      & 
      &   & \mbox{diag}(W) = 0, \ \mbox{diag}(Q) = 0. \; \\
    \end{aligned}
  }
\eeqa
Parameter $\beta_1$ controls how much $W$ and $Q$ are allowed to be different from each other. 

In \cite{r14}, item reviews in the form of short texts were used as side information in the models described above. These models were shown to outperform the \SLIM method without side information, as well as other approaches that use side information, in the top-$N$ recommendation task.

\begin{figure}[t]
    \centering
    \includegraphics[width = \textwidth]{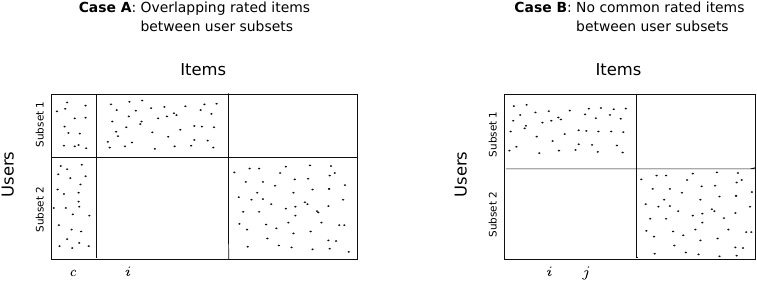}
    \caption[GLSLIM Example]{\GLSLIM Motivating Example. The figure shows the training matrices $R$ of two different datasets. Both  contain two user subsets. Let's assume that we are trying to compute recommendation scores for item $i$, and that the recommendations are computed using an item-item similarity-based method. Observe that in Case A there exist a set of items that have been rated solely by users that belong to Subset 1, while also a set of items which have been rated by users in both Subsets. Notice that the similarities of items $c$ and $i$ will be different when estimated based on the feedback of (a) Subset 1 alone; (b) Subset 2 alone; and, (c) the complete set of users. Specifically, their
similarity will be zero for the users of Subset 2 (as item $i$ is not rated by the users in that subset), but it will be e.g., $l_{ic} > 0$ for the users of Subset 1, as well as e.g., $g_{ic}  > 0$ when estimated globally, with $g_{ic}$ being potentially different than the locally estimated $l_{ic}$. 
Combining global and local item-item similarity models, in settings like this could help capture potentially diverse user preferences which would otherwise be missed if only a single global model, was computed instead. On the other hand, for datasets like the one pictured in Case B the similarity between e.g., items $i$ and $j$ will be the same, regardless of whether it is estimated globally, or locally for Subset 1, since both items have been rated only by users of Subset 1. }
    \label{fig:GLSLIM}
\end{figure}

\paragraph{\textbf{Global and local sparse neighborhood models}}
A global  item model may not always be sufficient to capture the preferences of every user; especially when there exist subsets of users with diverse or even opposing preferences. In cases like these training \textit{local item models} (i.e., item models that are estimated based on user subsets) is expected to be beneficial compared to adopting a single item model for all users in the system. An example of such a case can be seen in Figure~\ref{fig:GLSLIM}.

\GLSLIM~\cite{GLSLIM} aims to address the above issue. In a nutshell, \GLSLIM computes top-$N$ recommendations that utilize user-subset specific models (local models) and a global model. These models are jointly optimized along with computing the user-specific parameters that weigh their contribution in the production of the final recommendations. The underlying model used for the estimation of local and global item similarities is \SLIM.

Specifically, \GLSLIM estimates a global item-item coefficient matrix $S$ and also $k$ local item-item coefficient matrices $S^{p_{u}}$, where $k$ is the number of user subsets and $p_{u} \in\{1, \ldots, k\}$ is the index of the user subset, for which a local matrix $S^{p_{u}}$ is estimated. The recommendation score of user $u$, who belongs to subset $p_{u}$, for item $i$ is estimated as:
\begin{equation}
    \label{eq:GLSLIM_prediction}
    \tilde{r}_{ui} \, = \, \sum_{l \in R_{u}} g_{u} s_{li} \, + \, \left(1-g_u\right) s_{li}^{p_u}.
\end{equation}
Term $s_{li}$ depicts the global item-item similarity between the $l$-th item rated by $u$ and the target item $i$. Term $s_{li}^{p_{u}}$ captures the item-item similarity between the $l$-th item rated by $u$ and target item $i$, based on the local model that corresponds to user-subset, $p_{u}$, to which user $u$ belongs. Finally, the term $g_{u}\in[0,1]$ is the personalized weight of user $u$, which controls the involvement of the global and the local components, in the final recommendation score. 

The estimation of the global and the local item models, the user assignments to subsets, and the personalized weights is achieved by alternating minimization. Initially, the users are separated into subsets, using a clustering algorithm. Weights $g_{u}$ are initialized to 0.5 for all users, in order to enforce equal contribution of the global and the local components. Then the coefficient matrices $S$ and $S^{p_u},$ with $p_{u} \in\{1, \ldots, k\}$, as well as personalized weights $g_{u}$ are estimated, by repeating the following two step procedure: 

\vspace{2mm}
    \noindent\textbf{Step 1: Estimating local and global models:}  The training matrix $R$ is split into $k$ training matrices $R^{p_{u}}$ of size $|\set{U}| \times |\set{I}|,$ with $p_{u} \in\{1, \ldots, k\}$. Every row $u$ of $R^{p_{u}}$ coincides with the $u$-th row of $R$, if user $u$ belongs in the $p_{u}$-th subset; or is left empty, otherwise. In order to estimate the $i$-th column, $\sss_{i}$, of matrix $S$, and the $i$-th columns, $\sss_i^{p_{u}}$, of matrices $S^{p_u}, p_u \in\{1, \ldots, k\}$, \GLSLIM solves the following optimization problem:
     
    \begin{equation}
    \MINtwo{\sss_{i},\,\left\{\sss_{i}^{1}, \ldots, \sss_{i}^{k}\right\}}{\frac{1}{2}\left\|\rr_{i}-{\rm g} \odot R \sss_{i}-{\rm g}^{\prime} \odot \sum_{p_{u}=1}^{k} R^{p_{u}} \sss_{i}^{p_{u}}\right\|_{2}^{2} 
    \, + \, \frac{1}{2} \beta_{g}\left\|\sss_{i}\right\|_{2}^{2} 
    \, + \, \lambda_{g}\left\|\sss_{i}\right\|_{1}+ \\ 
    & + \, \sum_{p_{u}=1}^{k} \frac{1}{2} \beta_{l}\left\|\sss_{i}^{p_{u}}\right\|_{2}^{2} \, + \, \lambda_{l}\left\|\sss_{i}^{p_{u}}\right\|_{1} \\ }
    {\sss_{i} \geq 0, \ \ \sss_{i}^{p_{u}} \geq 0, \ \forall p_{u} \in\{1, \ldots, k\}}{[\sss_{i}]_i = 0, \ \ [\sss^{p_u}_i]_i=0, \ \forall p_{u}}
    \end{equation}

\noindent where $\rr_{i}$ is the $i$-th column of $R$; and,   $\beta_{g}$,  $\beta_{l}$ are the $l_{2}$ regularization hyperparameters corresponding to $S$,  $S^{p_{u}},  \forall p_{u} \in$ $\{1, \ldots, k\}$, respectively. Finally, $\lambda_{g}$, $\lambda_{l}$ are the $l_{1}$ regularization hyperparameters controlling the sparsity of $S$,  $S^{p_{u}}$ $\forall p_{u} \in\{1, \ldots, k\}$, respectively. The constraint $[\sss_{i}]_i=0$ makes sure that when computing $r_{ui}$, the element $r_{ui}$ is not used. Similarly, the constraints $[\sss^{p_u}_i]_i=0$ $\forall p_{u} \in\{1, \ldots, k\}$, enforce this property for the local sparse coefficient matrices as well. 

\vspace{2mm}
\noindent\textbf{Step 2: Updating user subsets:} 
With the global and local models fixed, \GLSLIM proceeds to update the user subsets. While doing that, it also determines the personalized weight $g_{u}$. Specifically, the computation of the personalized weight $g_{u}$, relies on minimizing the squared error of Equation~\eqref{eq:GLSLIM_prediction} for user $u$ who belongs to subset $p_{u}$, over all items $i$. Setting the derivative of the squared error to $0$, yields:
\begin{equation}
g_{u} \, = \, \frac{\sum_{i=1}^{m}\left(\sum_{l \in  R_u } s_{l i}-\sum_{l \in  R_u } s_{l i}^{p_{u}}\right)\left(r_{u i}-\sum_{l \in  R_u } s_{l i}^{p_{u}}\right)}{\sum_{i=1}^{m}\left(\sum_{l \in  R_u } s_{l i}-\sum_{l \in  R_u } s_{l i}^{p_{u}}\right)^{2}}.
\end{equation}
\GLSLIM tries to assign each user $u$ to every possible subset, while computing the weight $g_{u}$ that the user would have, if assigned to that subset. Then, for every subset $p_{u}$ and user $u$, the training error is computed and the user is assigned to the subset for which this error is minimized (or remains to the original subset, if no difference in training error occurs). 

Steps 1 and 2, are repeated until the number of users who switch subsets, in Step 2, becomes smaller than $1\%$ of $|\set{U}|$. It is empirically observed that initializing subset assignments with the $\mathtt{CLUTO}$~\cite{karypis2002cluto} clustering algorithm, results in a significant reduction of the number of iterations till convergence.    

Furthermore, a comprehensive set of experiments conducted in~\cite{GLSLIM} explore in detail the qualitative performance of \GLSLIM, and suggest that it improves upon the standard \SLIM, in several datasets. 

\subsection{Graph-based methods}

\begin{figure}[hbt]
\begin{center}
\scalebox{0.35}{\includegraphics{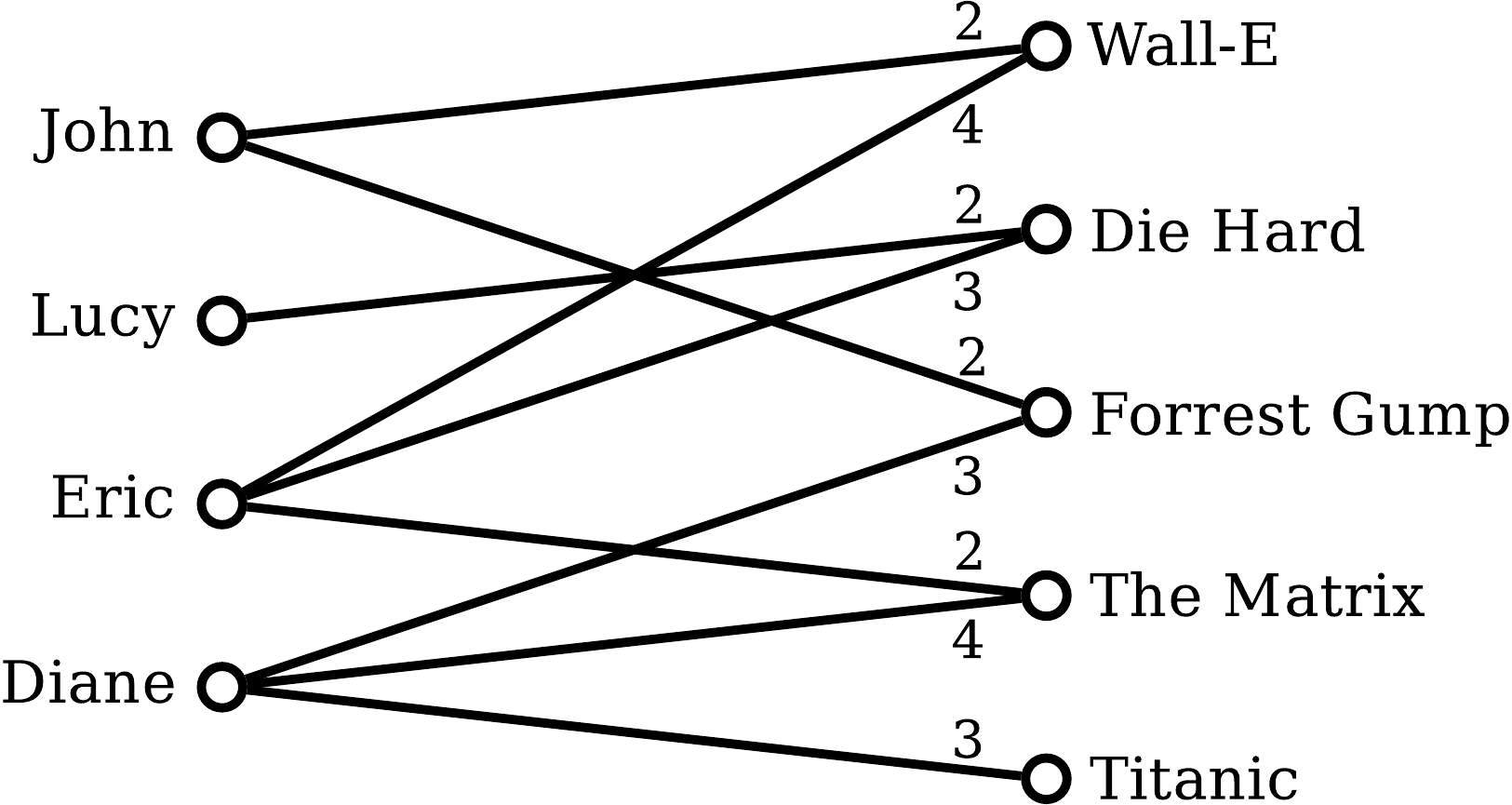}}
\end{center}
\caption{A bipartite graph representation of the ratings of Figure \ref{fig:toy-example} (\emph{only ratings with value in $\{2,3,4\}$ are shown}).}
\label{fig:bipartite}
\end{figure}

In graph-based approaches, the data is represented in the form of a graph where nodes are users, items or both, and edges encode the interactions or similarities between the users and items. For example, in Figure \ref{fig:bipartite}, the data is modeled as a bipartite graph where the two sets of nodes represent users and items, and an edge connects user $u$ to item $i$ if there is a rating given to $i$ by $u$ in the system. A weight can also be given to this edge, such as the value of its corresponding rating. In another model, the nodes can represent either users or items, and an edge connects two nodes if the ratings corresponding two these nodes are sufficiently correlated. The weight of this edge can be the corresponding correlation value.

In these models, standard approaches based on correlation predict the rating of a user $u$ for an item $i$ using only the nodes directly connected to $u$ or $i$. Graph-based approaches, on the other hand, allow nodes that are not directly connected to influence each other by propagating information along the edges of the graph. The greater the weight of an edge, the more information is allowed to pass through it. Also, the influence of a node on another should be less if the two nodes are further away in the graph. These two properties, known as \emph{propagation} and \emph{attenuation} \cite{gori07,huang04}, are often observed in graph-based similarity measures.

The transitive associations captured by graph-based methods can be used to recommend items in two different ways. In the first approach, the proximity of a user $u$ to an item $i$ in the graph is used directly to evaluate the relevance of $i$ to $u$ \cite{fouss07,gori07,huang04}. Following this idea, the items recommended to $u$ by the system are those that are the ``closest'' to $u$ in the graph. On the other hand, the second approach considers the proximity of two users or item nodes in the graph as a measure of similarity, and uses this similarity as the weights $w_{uv}$ or $w_{ij}$ of a neighborhood-based recommendation method \cite{fouss07,luo08}.

\subsubsection{Path-based similarity}

In path-based similarity, the distance between two nodes of the graph is evaluated as a function of the number and of paths connecting the two nodes, as well as the length of these paths. 

Let $R$ be once again the $|U|\!\times\!|I|$ rating matrix, where $r_{ui}$ is the rating given by user $u$ to an item $i$. The adjacency matrix $A$ of the user-item bipartite graph can be defined from $R$ as
\begin{displaymath}
 A \, = \, \left( \begin{array}{cc}
              0 & \, \tr{R} \\
              R & 0   \\
              \end{array} \right).
\end{displaymath}
The association between a user $u$ and an item $i$ can be defined as the sum of the weights of all distinctive paths connecting $u$ to $v$ (allowing nodes to appear more than once in the path), whose length is no more than a given maximum length $K$. Note that, since the graph is bipartite, $K$ should be an odd number. In order to attenuate the contribution of longer paths, the weight given to a path of length $k$ is defined as $\alpha^k$, where $\alpha \in [0,1]$. Using the fact that the number of length $k$ paths between pairs of nodes is given by $A^k$, the user-item association matrix $S_K$ is
\begin{eqnarray}
  S_K & \, = \, & \ssum_{k = 1}^K \alpha^k A^k \nonumber\\
      & \, = \, & (I - \alpha A)^{-1}(\alpha A - \alpha^K A^K).
\end{eqnarray}

This method of computing distances between nodes in a graph is known as the \emph{Katz} measure \cite{katz53}. Note that this measure is closely related to the \emph{Von Neumann Diffusion} kernel \cite{fouss06,kondor02,kunegis08}
\begin{eqnarray}
  K_\mathrm{VND} & \, = \, & \ssum_{k = 0}^\infty \alpha^k A^k \nonumber\\
                 & \, = \, & (I - \alpha A)^{-1}
\end{eqnarray}
and the \emph{Exponential Diffusion} kernel
\begin{eqnarray}
  K_\mathrm{ED} & \, = \, & \ssum_{k = 0}^\infty \frac{1}{k!} \alpha^k A^k \nonumber\\
                & \, = \, & \exp(\alpha A),
\end{eqnarray}
where $A^0 = I$.

In recommender systems that have a large number of users and items, computing these association values may require extensive computational resources. In \cite{huang04}, spreading activation techniques are used to overcome these limitations. Essentially, such techniques work by first activating a selected subset of nodes as starting nodes, and then iteratively activating the nodes that can be reached directly from the nodes that are already active, until a convergence criterion is met.

Path-based methods, as well as the other graph-based approaches described in this section, focus on finding relevant associations between users and items, not predicting exact ratings. Therefore, such methods are better suited for item retrieval tasks, where explicit ratings are often unavailable and the goal is to obtain a short list of relevant items (i.e., the top-$N$ recommendation problem). 

\subsubsection{Random walk similarity}

Transitive associations in graph-based methods can also be defined within a probabilistic framework. In this framework, the similarity or affinity between users or items is evaluated as a probability of reaching these nodes in a random walk. Formally, this can be described with a first-order Markov process defined by a set of $n$ states and a $n\!\times\!n$ transition probability matrix $P$ such that the probability of jumping from state $i$ to $j$ at any time-step $t$ is
\begin{displaymath}
  p_{ij} \, = \, \PP\big(s(t\!+\!1) = j \, | \, s(t) = i\big).
\end{displaymath}
Denote $\ppi(t)$ the vector containing the state probability distribution of step $t$, such that $\pi_i(t) = \PP\left(s(t) = i\right)$, the evolution of the Markov chain is characterized by 
\begin{displaymath}
  \ppi(t\!+\!1) \, = \, \tr{P} \ppi(t). 
\end{displaymath}
Moreover, under the condition that $P$ is row-stochastic, i.e. $\sum_j p_{ij} = 1$ for all $i$, the process converges to a stable distribution vector $\ppi(\infty)$ corresponding to the positive eigenvector of $\tr{P}$ with an eigenvalue of $1$. This process is often described in the form of a weighted graph having a node for each state, and where the probability of jumping from a node to an adjacent node is given by the weight of the edge connecting these nodes.

\paragraph{\bf Itemrank}

A recommendation approach, based on the PageRank algorithm for ranking Web pages \cite{brin98}, is ItemRank \cite{gori07}. This approach ranks the preferences of a user $u$ for unseen items $i$ as the probability of $u$ to visit $i$ in a random walk of a graph in which nodes correspond to the items of the system, and edges connects items that have been rated by common users. The edge weights are given by the $|\II|\!\times\!|\II|$ transition probability matrix $P$ for which $p_{ij} = |\UU_{ij}|/ |\UU_i|$ is the estimated conditional probability of a user to rate and item $j$ if it has rated an item $i$.

As in PageRank, the random walk can, at any step $t$, either jump using $P$ to an adjacent node with fixed probability $\alpha$, or ``teleport'' to any node with probability $(1-\alpha)$. Let $\rr_u$ be the $u$-th row of the rating matrix $R$, the probability distribution of user $u$ to teleport to other nodes is given by vector $\dd_u = \rr_u / ||\rr_u||$. Following these definitions, the state probability distribution vector of user $u$ at step $t\!+\!1$ can be expressed recursively as
\beq\label{eqn:itemrank}
  \ppi_u(t\!+\!1) \, = \, \alpha\tr{P}\ppi_u(t) \, + \, (1\!-\!\alpha) \dd_u.
\eeq
For practical reasons, $\ppi_u(\infty)$ is usually obtained with a procedure that first initializes the distribution as uniform, i.e. $\ppi_u(0) = \frac{1}{n} \ones$, and then iteratively updates $\ppi_u$, using (\ref{eqn:itemrank}), until convergence. Once $\ppi_u(\infty)$ has been computed, the system recommends to $u$ the item $i$ for which $\ppi_{ui}$ is the highest.

\paragraph{\bf Average first-passage/commute time}

Other distance measures based on random walks have been proposed for the recommendation problem. Among these are the \emph{average first-passage time} and the \emph{average commute time} \cite{fouss07,fouss06}.
The average first-passage time $m(j|i)$ \cite{norris99} is the average number of steps needed by a random walker to reach a node $j$ for the first time, when starting from a node $i \neq j$. Let $P$ be the $n\!\times\!n$ transition probability matrix, $m(j|i)$ can be obtained expressed recursively as
\begin{displaymath}
  m(j\,|\,i) \, = \, \left\{ \begin{array}{lll}
          0                               &, \ \ \textrm{if } i = j  \\
          1 + \ssum_{k=1}^n p_{ik} \, m(j\,|\,k) &, \ \ \textrm{otherwise}
          \end{array}\right.
\end{displaymath}
A problem with the average first-passage time is that it is not symmetric. A related measure that does not have this problem is the average commute time $n(i,j) = m(j\,|\,i)+m(i\,|\,j)$ \cite{gobel74}, corresponding to the average number of steps required by a random walker starting at node $i \neq j$ to reach node $j$ for the first time and go back to $i$. This measure has several interesting properties. Namely, it is a true distance measure in some Euclidean space \cite{gobel74}, and is closely related to the well-known property of resistance in electrical networks and to the pseudo-inverse of the graph Laplacian matrix \cite{fouss07}. 

In \cite{fouss07}, the average commute time is used to compute the distance between the nodes of a bipartite graph representing the interactions of users and items in a recommender system. For each user $u$ there is a directed edge from $u$ to every item $i \in \II_u$, and the weight of this edge is simply $1/|\II_u|$. Likewise, there is a directed edge from each item $i$ to every user $u \in \UU_i$, with weight $1/|\UU_i|$. Average commute times can be used in two different ways: 1) recommending to $u$ the item $i$ for which $n(u,i)$ is the smallest, or 2) finding the users nearest to $u$, according to the commute time distance, and then suggest to $u$ the item most liked by these users.

\subsubsection{{Combining Random Walks and Neighborhood-learning Methods}}

\paragraph{\textbf{Motivation and Challenges}}
Neighborhood-learning methods have been shown to achieve high top-$n$ recommendation accuracy while being scalable and easy to interpret. The fact, however, that they typically consider only direct item-to-item relations imposes  limitations to their quality and makes them brittle to the presence of sparsity, leading to poor itemspace coverage and substantial decay in performance. 
A promising direction towards ameliorating such problems involves treating item models as graphs onto which random-walk-based techniques can then be applied. However directly applying random walks on item models can lead to a number of problems that arise from their inherent mathematical properties and the way these properties relate to the underlying top-$n$ recommendation task. 

In particular, 
imagine of a random walker jumping from node to node on an item-to-item graph with  transition probabilities proportional to the proximity scores depicted by an item model $W$. If the starting distribution of this walker reflects the items consumed by a particular user $u$ in the past, the probability the walker  lands on different nodes after $K$ steps provide an intuitive measure of proximity that can be used to rank the nodes and recommend items to user $u$ accordingly. 

Concretely, if we denote the transition probability matrix of the walk $S = \operatorname{diag}(W \mathbf{1})^{-1} W$ where $\mathbf{1}$ is used to denote the vector of ones, personalized recommendations for user $u$ can be produced e.g., by leveraging the $K$-step landing distribution of a walk rooted on the items consumed by $u$;  
\begin{equation}
\piup_u^\top \, = \, \phiup_u^\top S^K, \qquad \phiup_u^\top \, = \, \tfrac{\rr_u^\top}{\lVert \rr_u^\top\rVert_1}
\label{model:srw}
\end{equation}
or by computing the limiting distribution of a random walk with restarts on $S$, using $\phiup_u^\top$ as the restarting distribution.  The latter approach is the well-known personalized PageRank model~\cite{brin98} with teleportation vector $\phiup_u^\top$ and damping factor $p$, and its stationary distribution can be expressed~\cite{langville2011google} as 
\begin{equation}
\piup_u^\top \, = \, \phiup_u^\top \sum_{k=0}^{\infty}(1-p) p^k S^k.
\label{model:simplepr}
\end{equation}   
Clearly, both schemes harvest the information captured in the $K$-step landing probabilities  $\{\phiup_u^\top S^k\}_{k=0,1,\dots}$. 
But, how do these landing probabilities behave as the number of steps $K$ increases? For how long will they still be significantly influenced by user's preferences $\phiup^\top_u$? 

Markov chain theory ensures that when $S$ is irreducible and aperiodic the landing probabilities will converge to a \textit{unique} stationary distribution irrespectively of the initialization of the walk. This means that for large enough $K$, the $K$-step landing probabilities will no longer be ``personalized,'' in the sense that they will become independent of the user-specific starting vector $\phiup_u^\top$.  Furthermore, long before reaching equilibrium, the quality of these vectors in terms of recommendation will start to plummet as more and more probability mass gets concentrated to the central nodes of the graph. Note, that the same issue arises for simple random walks that act directly on the user-item bipartite network, and has lead to methods that typically consider only very short-length random walks, and need to explicitly re-rank the $K$-step landing probabilities, in order to compensate for the inherent bias of the walk towards popular items~\cite{RP3b}. However, longer random-walks might be necessary to capture non-trivial multi-hop relations between the items, as well as to ensure  better coverage of the itemspace.

\paragraph{\textbf{The RecWalk Recommendation Framework}}
\texttt{RecWalk}~\cite{10.1145/3289600.3291016,10.1145/3406241} addresses the aforementioned challenges, and resolves this long- vs short-length walk dilemma through the construction of a \textit{nearly uncoupled random walk}~\cite{NCD1,NCD2} that gives full control over the stochastic dynamics of the walk towards equilibrium; provably, and irrespectively of the dataset or the specific item model onto which it is applied. Intuitively, this allows for prolonged and effective exploration of the underlying network while keeping the influence of the user-specific initialization strong.\footnote{The mathematical details behind the particular construction choices of \recwalk that enforce such desired mixing properties can be found in~\cite{10.1145/3289600.3291016,10.1145/3406241}. } 

\begin{figure}[t]
    \centering
 \includegraphics[angle=0,origin=c]{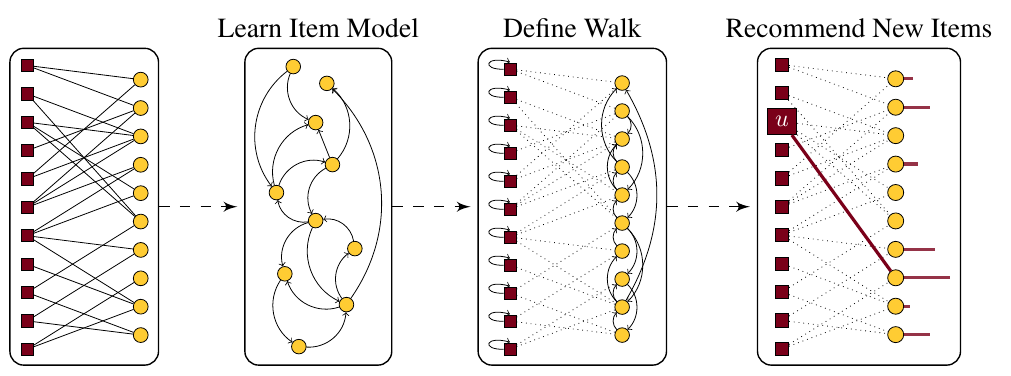}
    \caption{RecWalk Illustration. Maroon colored nodes correspond to users; Gold colored nodes correspond to items. 
    }
    \label{fig:RecWalk}
\end{figure}

 From a random-walk point of view, the \recwalk model can be described as follows: Consider a random walker jumping from node to node on the user-item bipartite network. Suppose the walker currently occupies a node $c \in \set{U}\cup\set{I}$. In order to determine the next step transition the walker tosses a biased coin that yields heads with probability $\alpha$ and tails with probability $(1-\alpha)$: 
\begin{enumerate}
	\item If the coin-toss yields \textit{heads}, then: 
	\begin{enumerate}
		\item if $c \in \set{U}$, the walker jumps to one of the items rated by the current user (i.e., the user corresponding to the current node $c$) uniformly at random;
		\item if $c \in \set{I}$, the walker jumps to one of the users that have rated the current item uniformly at random; 
	\end{enumerate} 
	\item If the coin-toss yields \textit{tails}, then:
	\begin{enumerate}
		\item if $c\in\set{U}$, the walker stays put;
		\item if $c \in \set{I}$, the walker jumps to a related item abiding by an \textit{item-to-item transition probability matrix} $M_\set{I}$, that is defined in terms of an underlying item model.
	\end{enumerate} 
\end{enumerate}   
The stochastic process that describes this random walk is defined to be a homogeneous discrete time Markov chain with state space $\set{U}\cup\set{I}$; i.e., the transition probabilities from any given node $c$ to the other nodes, are fixed and independent of the nodes visited by the random walker before reaching $c$. An illustration of the \recwalk model is given in Figure~\ref{fig:RecWalk}. 

The transition probability matrix $P$ that governs the behavior of the random walker can be usefully expressed as a weighted sum of two stochastic matrices $H$ and $M$ as
\begin{equation}
P \, = \, \alpha H \, + \, (1-\alpha) M \label{transitionProbabilityMatrixP}
\end{equation} 
where $ 0< \alpha < 1 $, is a parameter that controls the involvement of these two components in the final model.
Matrix $H$ can be thought of as the transition probability matrix of a simple random walk on the user-item bipartite network. Assuming that the rating matrix $R$ has no zero columns and rows, matrix $H$ can be expressed as   
\begin{equation}
H  \, = \, \Diag(A\ones)^{-1}A, \qquad \textrm{where } \   A \, = \, \left( \begin{array}{cc}
                & \, R \\
              \tr{R} &     \\
              \end{array} \right).
\end{equation} 
Matrix $M$, is defined as 
\begin{equation}
M \, = \, \pmat{ I &  \\ 
	 & M_\set{I} }
\end{equation}
where $I \in \mathfrak{R}^{U \times U}$ the identity matrix and  $M_\set{I} \in \mathfrak{R}^{I \times I}$ is a transition probability matrix designed to capture relations between the items. In particular, given an item model with non-negative weights $W$ (e.g., the aggregation matrix produced by a \SLIM model), matrix $M_\set{I}$ is defined using the following stochasticity adjustment strategy: 
\begin{equation}
M_\set{I} \, = \, \frac{1}{\lVert W \rVert_\infty} W \, + \, \Diag\left(\ones-\frac{1}{\lVert W \rVert_\infty}W\ones\right).
\label{def:M_I}
\end{equation}   
The first term divides all the elements by the maximum row-sum of $W$ and the second enforces stochasticity by adding residuals to the diagonal, appropriately. The motivation behind this definition is to retain the information captured by the relative differences of the item-to-item relations in $W$. This prevents items that are loosely related to the rest of the itemspace to disproportionately influence the inter-item transitions and introduce noise to the model.\footnote{From a purely mathematical point-of-view the above strategy promotes desired spectral properties to $M_\set{I}$ that are shown to be intertwined with recommendation performance. For additional details see~\cite{10.1145/3406241}.} 
 
In \recwalk the recommendations are produced by exploiting the information captured in the successive landing probability distributions of a walk initialized in a user-specific way. 
Two simple recommendation strategies that were considered in~\cite{10.1145/3289600.3291016} are:   
\begin{description}
	\item[\recwalkKstep:] The recommendation score of user $u$ for item $i$ is defined to be the probability the random walker lands on node $i$ after $K$ steps, given that the starting node was $u$. In other words, the recommendation score for item $i$ is given by the corresponding elements of   
	\begin{equation}
	\label{eq:recwalk_k_step}
	\piup_u^\top \, = \, \eee_u^\top P^K
	\end{equation} 
	where $\eee_u \in \mathfrak{R}^{U+I}$ is a vector that contains the element 1 on the position that corresponds to user $u$ and zeros elsewhere. The computation of the recommendations is performed by $K$ sparse-matrix-vector products with matrix $P$, and it entails
	\begin{math}
	\upTheta(K\operatorname{nnz}(P))
	\end{math}
	operations, where $\operatorname{nnz}(P)$ is the number of nonzero elements in $P$. 
	\item[\recwalkPR:] The recommendation score of user $u$ for item $i$ is defined to be the element that corresponds to item $i$ in the limiting distribution of a random walk with restarts on $P$, with restarting probability $\eta$ and restarting distribution $\eee_u$:  
	\begin{equation}
	\label{eq:recwalk_pr}
	\piup_u^\top \, = \, \lim\limits_{K\to \infty}\, \eee_u^\top\big(\eta P \, + \, (1-\eta)\ones \eee_u^\top\big)^K. 
	\end{equation} 
	The limiting distribution in \eqref{eq:recwalk_pr} can be computed efficiently using e.g., the power method, or any specialized PageRank solver. Note that this variant of \recwalk  also comes with theoretical guarantees for item-space coverage for every user in the system, regardless of the base item model $W$ used in the definition of matrix $M_\set{I}$~\cite{10.1145/3406241}. 
\end{description} 
	In \cite{10.1145/3406241} it was shown that both approaches manage to boost the quality of several base item models on top of which they were built. Using \fsSLIM~\cite{ning2011slim} with small number of neighbors as a base item model, in particular, was shown to achieve state-of-the-art recommendation performance, in several datasets. At the same time \recwalk was found to dramatically increase  itemspace coverage of the produced recommendations, in every considered setting. This was true both for \recwalkKstep, as well as for \recwalkPR. 
	
	

\subsubsection{User-Adaptive Diffusion Models} 


 \noindent{\textbf{Motivation}}: 
Personalization of the recommendation vectors in the graph-based schemes we have seen thus far, comes from the use of a user-specific initialization, or a user-specific restarting distribution. 
However, the underlying mechanism  for propagating user preferences, across the itemspace (i.e., the adopted diffusion function, or the choice of the $K$-step distribution) is \textit{fixed} for every user in the system. From a user modeling point of view this translates to the implicit assumption that every user explores the itemspace in \textit{exactly the same} way---overlooking the reality that different users can have different behavioral patterns. The fundamental premise of \perdif~\cite{PERDIF} is that the latent item exploration behavior of the users can be captured better by \textit{user-specific} preference propagation mechanisms; thus, leading to improved recommendations. 

\perdif proposes a simple model of personalized item exploration subject to an underlying item model. 
At each step the users might either decide to go forth and discover items related to the ones they are currently considering, or return to their base and possibly go down alternative paths. Different users, might explore the itemspace in different ways; and their behavior might change throughout the exploration session. The following stochastic process, formalizes the above idea:

\vspace{2mm}
\noindent{\textbf{The PerDIF Item Discovery Process:}} Consider a random walker carrying a bag of $K$ biased coins. The coins are labeled with consecutive integers from 1 to $K$. Initially, the random walker occupies the nodes of graph according to distribution $\phiup$. She then flips the 1st coin: if it turns heads (with probability $\mu_1$), she jumps to a different node in the graph abiding by the probability matrix $P$; if it turns tails (with probability $1-\mu_1$), she jumps to a node according to the probability distribution $\phiup$. She then flips the 2nd coin and she either follows $P$ with probability $\mu_2$ or `restarts' to $\phiup$ with probability ($1-\mu_2$). The walk continues until she has used all her $K$ coins.
At the $k$-th step the transitions of the random walker are completely determined by the probability the $k$-th coin turning heads ($\mu_k$), the transition matrix $P$, and the restarting distribution $\phiup$. Thus, the stochastic process that governs the position of the random walker over time is a time-inhomogeneous Markov chain with state space the nodes of the graph, and transition matrix at time $k$ given by
\begin{equation}
G(\mu_k) \, = \, \mu_k P \, + \, (1-\mu_k)\ones\phiup^\top.
\end{equation}
The node occupation distribution of the random walker after the last transition can therefore be expressed as
\begin{equation}
\piup^\top \, = \, \phiup^\top G(\mu_1)\,G(\mu_2)\,\cdots\, G(\mu_K).
\label{eq:itemExporationProcess}
\end{equation}

Given an item transition probability matrix $P$, and a user-specific restarting distribution $\phiup_u$, 
the goal is to find a set of probabilities $\muup_u= \pmat{\mu_1,\dots,\mu_K}$ so that the outcome of the aforementioned item exploration process yields a meaningful distribution over the items that can be used for recommendation. \perdif tackles this task as follows:

\vspace{2mm}

\noindent{\textbf{Learning the personalized probabilities:}} For each user $u$ we randomly sample one item she has interacted with (henceforth referred to as the `target' item) alongside $\tau_{\mathit{neg}}$ unseen items, and we fit $\muup_u$ so that the node occupancy distribution after a $K$-step item exploration process rooted on $\phiup_u$ (cf~\eqref{eq:itemExporationProcess}) yields high probability to the target item while keeping the probabilities of the negative items low. Concretely, upon defining a vector $\hh_u\in\mathfrak{R}^{\tau_{\mathit{neg}}+1}$ which contains the value 1 for the target item and zeros for the negative items, we learn $\muup_u$ by solving
\begin{equation}
\MINone{\muup_u\in \mathfrak{R}^K}{ \big\lVert \phiup_u^\top G(\mu_1)\cdots G(\mu_K)E_u - \hh_u^\top \big\rVert_2^2}{\mu_i \in (0,1), \quad \forall i \in [1,\dots,K]}
\label{problem1}
\end{equation}
where $\mu_i = [\muup_u]_i, \forall i$, and $E_u$ is a $(I\times (\tau_{\mathit{neg}}+1))$  matrix designed to select and rearrange the elements of the vector $\phiup_u^\top G(\mu_1)\cdots G(\mu_K)$ according to the sequence of items comprising $\hh_u$. Upon obtaining $\muup_u$, personalized recommendations for user $u$ can be computed as
\begin{equation}\label{eq:recommendationsMuForm}
\piup_u^\top = \phiup_u^\top G(\mu_1)\cdots G(\mu_K).
\end{equation}
Leveraging the special properties of the stochastic matrix $G$ the above non-linear optimization problem can be solved efficiently. In particular, it can be shown~\cite{PERDIF} that the optimization problem~\eqref{problem1} is equivalent to 
	\begin{displaymath}
	\MIN{\omegaup_u \in \Delta_{++}^{K+1}}{\, \big\lVert \omegaup_u^\top S_u E_u - \hh_u^\top \big\rVert_2^2}
	\end{displaymath}
	where $\Delta_{++}^{K+1} = \{x: x^\top\ones = 1, x>0 \}$ and
	\begin{displaymath}
	S_u \, = \,
	\pmat{
		\phiup_u^\top \\
		\phiup_u^\top P \\
		\phiup_u^\top P^2 \\
		\vdots \\[0.1cm]
		\phiup_u^\top P^K
	}, \qquad
	\omegaup_u \, \equiv \, \omegaup_u(\muup_u) \, = \, \pmat{
		1-\mu_K\\
		\mu_K\,(1-\mu_{K-1})\\
		\mu_K\,\mu_{K-1}\,(1-\mu_{K-2})\\
		\vdots\\
		\mu_K\,\cdots\,\mu_2\,(1-\mu_1)\\
		\mu_K\,\cdots\,\mu_2\,\mu_1
	}.
	\end{displaymath}

The above result simplifies learning $\muup_u$ significantly. It also lends \perdif its name. In particular, the task of finding personalized probabilities for the item exploration process,  reduces to that of finding \textit{personalized diffusion coefficients} $\omegaup_u$ over the space of the first $K$ landing probabilities of a walk rooted on $\phiup_u$ (see definition of $S_u$). Afterwards $\muup_u$ can be obtained in linear time from $\omegaup_u$ upon solving a simple forward recurrence~\cite{PERDIF}. Taking into account the fact that in recommendation settings $K$ will typically be small and $\phiup_u, P$ sparse,  building `on-the-fly' $S_u E_u$ row-by-row, and solving the $(K+1)$-dimensional convex quadratic problem
\begin{equation}\label{eq:free_learn}
	\free\,: \ \ \MIN{\omegaup_u \in \Delta_{++}^{K+1}}{ \big\lVert \omegaup_u^\top S_u E_u - \hh_u^\top \big\rVert_2^2}
\end{equation}
can be performed very efficiently (typically in a matter of milliseconds  even in large scale settings).

Moreover, working on the space of landing probabilities can also facilitate parametrising the diffusion coefficients within a  family of known diffusions. This motivates the parameterized variant of \perdif
\begin{equation}\label{eq:dict_learn}
	\dict:\ \ \MIN{\gammaup_u \in \Delta_{+}^{L}}{ \lVert \gammaup_u^\top D S_u E_u - \hh_u^\top \rVert_2^2}
\end{equation}
with $\Delta_{+}^{L}=\{y:y^\top\ones = 1, y\geq0\}$ and $D \in \mathfrak{R}^{L\times (K+1)}$ defined such that its rows contain preselected diffusion coefficients (e.g., PageRank~\cite{brin98} coefficients for several damping factors, heat kernel~\cite{chung2007heat} coefficients for several \textit{temperature} values etc.), normalized to sum to one. Upon obtaining $\gammaup_u$, vector $\omegaup_u$ can be computed as $\omegaup_u^\top = \gammaup_u^\top D$.  

\begin{figure}[t]
    \centering
    \includegraphics[scale=1.19]{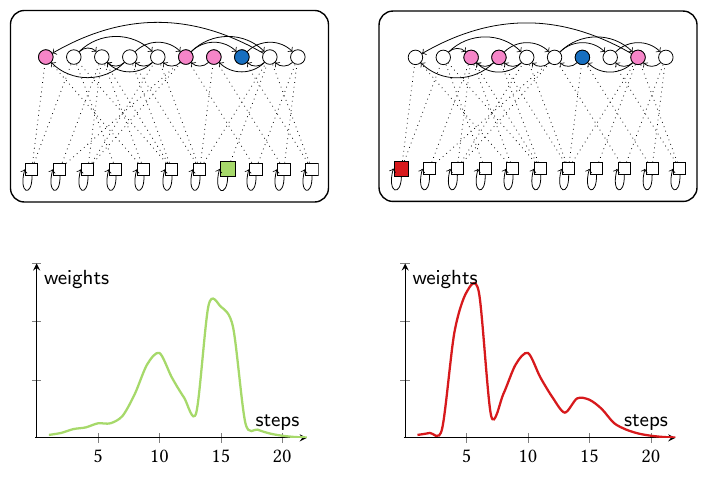}
    \caption{Personalized Diffusions on the User-Item Bipartite Network. }
    \label{fig:PerDif}
\end{figure}
While \free learns $\omegaup_u$ by weighing the contributions of the landing probabilities directly, \dict constrains $\omegaup_u$ to comprise a user-specific mixture of predetermined such weights (i.e., the rows of $D$), thus allowing one to endow $\omegaup_u$ with desired properties, relevant to the specific recommendation task at hand. 
Furthermore, the use of matrix $D$ can improve the robustness of the personalized diffusions in settings where the recommendation quality of the individual landing distributions comprising $S_u$ is uneven across the $K$ steps considered.

Besides, its merits in terms of recommendation accuracy, personalizing the diffusions within the \perdif framework can also provide useful information arising from the analysis of the learned diffusion coefficients, $\omegaup_u$. In particular, the dual interpretation of the model parameters ($\muup_u$ in the item exploration space; and, $\omegaup_u$ in the diffusion space)  allows utilizing the learned model parameters to identify users for which the model will most likely lead to poor predictions, at training-time---thereby affording preemptive interventions to handle such cases appropriately. This affords a level of transparency that can prove particularly useful in practical settings (for the technical details on how this can be achieved see~\cite{PERDIF}). 

\section{Conclusion}

One of the earliest approaches proposed for the task item recommendation, neighbor\-hood-based recommendation still ranks among the most popular methods for this problem. Although quite simple to describe and implement, this recommendation approach has several important advantages, including its ability to explain a recommendation with the list of the neighbors used, its computational and space efficiency which allows it to scale to large recommender systems, and its marked stability in an online setting where new users and items are constantly added. Another of its strengths is its potential to make serendipitous recommendations that can lead users to the discovery of unexpected, yet very interesting items.  

In the implementation of a neighborhood-based approach, one has to make several important decisions. Perhaps the one having the greatest impact on the accuracy and efficiency of the recommender system is choosing between a user-based and an item-based neighborhood method. In typical commercial recommender systems, where the number of users far exceeds the number of available items, item-based approaches are typically preferred since they provide more accurate recommendations, while being more computationally efficient and requiring less frequent updates. On the other hand, user-based methods usually provide more original recommendations, which may lead users to a more satisfying experience. Moreover, the different components of a neighborhood-based method, which include the normalization of ratings, the computation of the similarity weights and the selection of the nearest-neighbors, can also have a significant influence on the quality of the recommender system. For each of these components, several different alternatives are available. Although the merit of each of these has been described in this document and in the literature, it is important to remember that the ``best'' approach may differ from one recommendation setting to the next. Thus, it is important to evaluate them on data collected from the actual system, and in light of the particular needs of the application. 

Modern machine-learning-based techniques can be used to further increase the performance of neighborhood-based approaches, by automatically extracting the most representative neighborhoods based on the available data. Such models achieve state-of-the-art recommendation accuracy, however their adoption imposes additional computational burden that needs to be considered in light of the particular characteristics of the recommendation problem at hand.
Finally, when the performance of a neighborhood-based approach suffers from the problems of limited coverage and sparsity, one may explore techniques based on dimensionality reduction or graphs. Dimensionality reduction provides a compact representation of users and items that captures their most significant features. An advantage of such approach is that it allows to obtain meaningful relations between pairs of users or items, even though these users have rated different items, or these items were rated by different users. On the other hand, graph-based techniques exploit the transitive relations in the data. These techniques also avoid the problems of sparsity and limited coverage by evaluating the relationship between users or items that are not ``directly connected''. However, unlike dimensionality reduction, graph-based methods also preserve some of the ``local'' relations in the data, which are useful in making serendipitous recommendations.


\end{document}